\documentclass[prd,onecolumn,amsmath,amssymb,floatfix,nofootinbib]{revtex4}
\usepackage{amssymb}
\usepackage{amsmath}
\usepackage{feynmp}
\usepackage{slashed}

\usepackage{caption}
\usepackage{subfigure}

\usepackage{graphicx}
\usepackage{graphics}
\usepackage{dcolumn}
\usepackage{color}
\usepackage{rotate}
\usepackage{fancyhdr}
\usepackage{hyperref}
\hypersetup{
    colorlinks=true,
    linkcolor=black,
    filecolor=black,
    urlcolor=black,
}
\usepackage{indentfirst}

\usepackage{umoline}

\def\be{\begin{eqnarray}}
\def\ee{\end{eqnarray}}
\def\no{\nonumber}

\newcommand{\abs}[1]{\lvert#1\rvert}

\def\lsim{\mathrel{\rlap{\lower4pt\hbox{\hskip1pt$\sim$}}
     \raise1pt\hbox{$<$}}}         
\def\gsim{\mathrel{\rlap{\lower4pt\hbox{\hskip1pt$\sim$}}
     \raise1pt\hbox{$>$}}}         
\newcommand{\beq}{\begin{equation}}
\newcommand{\eeq}{\end{equation}}

\newenvironment{Eqnarray}{\arraycolsep 0.14em\begin{eqnarray}}{\end{eqnarray}}
\def\beqa{\begin{Eqnarray}}
\def\eeqa{\end{Eqnarray}}

\begin{document}
\title{Exotic colored scalars at the LHC
}

\author{Kfir Blum$^{1}$, Aielet Efrati$^{1}$, Claudia Frugiuele$^{1}$, Yosef Nir$^{1}$}
\affiliation{$^1$Department of Particle Physics and Astrophysics, Weizmann Institute of Science, Rehovot, Israel 7610001}
\email{$^a$kfir.blum, aielet.efrati, claudia.frugiuele, yosef.nir@weizmann.ac.il}

\begin{abstract}
We study the phenomenology of exotic color-triplet scalar particles $X$ with charge $|Q|=2/3, 4/3,5/3,7/3,8/3$ and $10/3$. If $X$ is an $SU(2)_W$-non-singlet, mass splitting within the multiplet allows for cascade decays of the members into the lightest state. We study examples where the lightest state, in turn, decays into a three-body $W^\pm jj$ final state, and show that in such case the entire multiplet is compatible with indirect precision tests and with direct collider searches for continuum pair production of $X$ down to $m_X\sim250$~GeV.
However, bound states $S$, made of $XX^\dag$ pairs at $m_S\approx2m_X$, form under rather generic conditions and their decay to diphoton can be the first discovery channel of the model.
Furthermore, for $SU(2)_W$-non-singlets, the mode $S\to W^+W^-$ may be observable and the width of $S\to\gamma\gamma$ and $S\to jj$ may appear large as a consequence of mass splittings within the $X$-multiplet. As an example we study in detail the case of an $SU(2)_W$-quartet, finding that $m_X\simeq450$~GeV is allowed by all current searches.
\end{abstract}

\maketitle

\tableofcontents

\section{Introduction}
The large hadron collider (LHC) search for new physics at or below the TeV scale is far from complete, even for strongly interacting particles. As concerns the commonly studied Standard Model (SM) extensions~\cite{Dimopoulos:1981zb,Kaplan:1983sm,Giudice:2013nak}, the dedicated searches by CMS and ATLAS for new strongly interacting light degrees of freedom are covering a large part of the parameter space.
However, new colored particles beyond these standard scenarios could  still have unexpected phenomenology and, in this case, traditional LHC searches often lose much of their power. In this work we consider colored scalar states with exotic EM charges, with a focus on $SU(2)_W$-non-singlets. Such particles, while being copiously produced at the LHC, could still be hiding undiscovered amidst the large QCD background. Three different paths can be pursued in the experimental search for these particles:
\begin{enumerate}
\item Direct collider searches for QCD continuum pair production of $X_Q$, a colored particle with EM charge $Q$. Such searches are potentially effective, but depend on the decay modes of $X_Q$ and hence are model dependent.
\item Precision measurements of electroweak (EW) processes, constituting an indirect search for $X_Q$.
\item Direct collider searches for $S_Q$, the bound state formed out of $X_QX^\dagger_Q$ through Coulomb gluon exchange, with mass $m_{S_Q}\simeq2m_{X_Q}$. $S_Q$ decays into diboson final states, with branching ratios that are determined to a large extent by the quantum numbers of $X_Q$. For exotic states the consequent constraints are often less model dependent than continuum pair production searches (see {\it e.g.}~\cite{Kats:2012ym,Kats:2016kuz}).
\end{enumerate}
We pursue all three avenues in this work.

The paper is organized as follows. In Sec.~\ref{sec:model} we present our theoretical framework and the relevant representations for our study. Sec.~\ref{sec:direct} details the experimental bounds from direct searches for continuum QCD pair production of $X_Q$. In Sec.~\ref{sec:cascade} we discuss mass splittings within $SU(2)_W$ multiplets and the implications for cascade decays. In Sec.~\ref{sec:cases} we present a benchmark model. Sec.~\ref{sec:SU2} deals with the unique phenomenology of $SU(2)_W$ multiplets, and the footprint it might leave in indirect probes such as electroweak precision measurements (EWPM), Higgs couplings and the renormalisation group evolution of various couplings. In Sec.~\ref{sec:bound} we study the QCD bound states  formed out of $X_QX_Q^\dagger$ pairs, and the possible signatures at the LHC. We conclude in Sec.~\ref{sec:concs}. Various technical details are presented in the Appendices.

\section{Theoretical framework}\label{sec:model}
Consider a scalar $X$ in the $(R,n)_Y$ representation of the $SU(3)_C\times SU(2)_W\times U(1)_Y$ gauge group.
The Lagrangian is given by
\beqa\label{eq:lagX}
\mathcal{L}=&&\mathcal{L}_{\rm SM}+\abs{D_\mu X}^2+\mathcal{L}_{Y_X}-V(H,X)\,,
\eeqa
where $D_\mu$ is the covariant derivative, determined by the quantum numbers (QN) of $X$, and $H$ is the SM Higgs doublet, $H\sim(1,2)_{+1/2}$. The scalar potential $V(H,X)$ has the form
\beqa\label{eq:Spotential}
V(H,X)=&&V^{\rm SM}(H)+m_X^2 X^\dagger X+\frac{\lambda_X}{2}(X^\dagger X)^2+\lambda_{XH}X^\dagger X H^\dagger H+\lambda^\prime_{XH}(X^\dagger T^a_n X)(H^\dagger T^a_2 H)\,,
\eeqa
where $T^a_n$ are the $SU(2)_W$ generators in the $n$ representation.
As we explore below, $\lambda^\prime_{XH}\neq0$ generates mass splitting between the various states $X_Q$. Both $\lambda_{XH}\neq0$ and $\lambda_{XH}^\prime\neq0$ modify Higgs couplings to SM fermions and gauge bosons.

We comment that Eq.~(\ref{eq:Spotential}) is not the most general form possible for $V(H,X)$.
Additional $X^4$ couplings may arise {\it e.g.} for color triplets in a non-singlet $SU(2)_W$ representation.
As long as these couplings are small compared to $g_s^2\sim1$, they are not essential in most of our analysis and we omit them here.

As concerns the $SU(3)_C$ representation of $X$, we focus on color-triplets. This is a common starting point in many analyses, often considering quantum numbers similar to those of the SM quarks as occurs in supersymmetric models. The common lore is that first and second generation squarks are ruled out below 1.4~TeV while stops should be heavier than 900~GeV~\cite{Craig:2015yvw}. We study how this discussion is affected once exotic $SU(2)_W\times U(1)_Y$ representations are considered.

Other $SU(3)_C$ assignments have been studied in various contexts. For instance, supersymmetric models with Dirac gauginos introduce a color-octet scalar as the superpartner of the fermion which marries the gluino to form a Dirac fermion~\cite{Fox:2002bu}. Color-sextets have been introduced in some models of grand unification~\cite{Pati:1974yy,Chacko:1998td}.
Despite this interest, we keep our focus on $R=3$ for concreteness, though we include generic representations $R$ in some parts of the analysis where it does not introduce excess clatter.

The terms in $\mathcal{L}_{Y_X}$ break $X$ number and thus control the decay of $X$ to SM final states. With some abuse of notation, we refer to the terms in $\mathcal{L}_{Y_X}$ as Yukawa interactions. We maintain this terminology also to nonrenormalizable operators which, when the Higgs fields are replaced by their vacuum expectation values, lead to effective Yukawa couplings of $X$ with SM fermions.
A doublet or a triplet of $SU(2)_W$ can couple to a fermion pair in a renormalizable operator, while  other representations of $SU(2)_W$ require higher dimensional operators for the decay of their members.
The inclusion of effective operators truncates the validity of our model at some cut-off scale $\Lambda$.
To avoid the need for low cut-off scale, we restrict our discussion to effective operators with mass dimension $\leq6$.
This, in turn, leads us to consider $n\leq5$, and limits the possible hypercharge assignments for $X$.

In Table~\ref{tab:reps} we list all possible representations of $X$, for which we can find $X$-decay operators compatible with the restriction $d\leq6$ for $\mathcal{L}_{Y_X}$. We also list the corresponding diquark and/or leptoquark $X$-number violating operators. We denote the SM left-handed doublets as $Q$ and $L$, and the right-handed singlets as $U,D$ and $E$. Throughout the analysis we will assume that, when several operators are available in Table~\ref{tab:reps}, only one of them exists while the others are absent or negligible. For brevity, we omit $d\leq6$ operators which include derivative interactions, as they introduce no new representations for $X$.
\begin{table}[h!]
\caption{ \it $X$ representations with the corresponding $d\leq6$ operators inducing decay of $X$ to SM final states.}
\label{tab:reps}
\begin{center}
\begin{tabular}{cccc} \hline\hline
\rule{0pt}{1.0em}%
$(R,n)_Y$ & $\abs{Q}_{\rm high}$ & Hadronic operators & Lepto-quark operators \\[2pt] \hline\hline
\rule{0pt}{1.0em}%
$(3,1)_Y$    & $|Y|$  & & \\ \hline \rule{0pt}{1.0em}
$(3,2)_{+1/6}$ & $2/3$
& $XDUH^\dagger$, $XD_{\left[i\right.}D_{\left.j\right]}H$, $XQ_{\left\{i\right.}Q_{\left.j\right\}}H^\dagger$ & $X\bar DL$, $X\bar U\bar EH^\dagger$, $X\bar Q\bar LH^\dagger$,  $ X\bar ULHH$, $X\bar QEHH$ \\ \hline\rule{0pt}{1.0em}
$(3,2)_{-5/6}$ & $4/3$
& $XQ_{\left\{i\right.}Q_{\left.j\right\}}H$, $XU_{\left[i\right.}U_{\left.j\right]}H^\dagger$, $XUDH$ & $X\bar Q\bar LH$, $X\bar U\bar EH$, $X\bar D\bar EH^\dagger$, $X\bar DLHH$ \\
$(3,3)_{-1/3}$ & $4/3$
& $XQ_{\left[i\right.}Q_{\left.j\right]}$, $XUDH^\dag H$, $XU_{\left[i\right.}U_{\left.j\right]}H^\dag H^\dag$, $XD_{\left[i\right.}D_{\left.j\right]}HH$ & $X\bar Q\bar L$ , $X\bar DLH$, $X\bar D\bar EH^\dag H^\dag$\\ \hline\rule{0pt}{1.0em}
$(3,2)_{+7/6}$ &  $5/3$
& $XD_{\left[i\right.}D_{\left.j\right]}H^\dagger$ & $X\bar QE$, $X\bar UL$, $XL\bar DH^\dag H^\dag$ \\
$(3,3)_{+2/3}$ & $5/3$
& $XD_{\left[i\right.}D_{\left.j\right]}HH^\dagger$, $XUDH^\dagger H^\dagger$ & $X\bar QEH$, $X\bar ULH$, $X\bar DLH^\dagger$  \\
{$(3,4)_{+1/6}$} & $5/3$
& $XQ_{\left[i\right.}Q_{\left.j\right]}H^\dagger$ & $X\bar Q\bar LH^\dagger$, $X\bar QEHH$, $X\bar DLH^\dagger H$, $X\bar ULHH$ \\ \hline\rule{0pt}{1.0em}
$(3,2)_{-11/6}$ & $7/3$
& $XU_{\left[i\right.}U_{\left.j\right]}H$ &  $X\bar D\bar EH$ \\
$(3,3)_{-4/3}$ & $7/3$ &
$XU_{\left[i\right.}U_{\left.j\right]}H^\dagger H$, $XUDHH$, $XQ_{\left\{i\right.}Q_{\left.j\right\}}H^\dagger H$  & $X\bar U\bar EHH$, $X\bar D\bar EH^\dagger H$ \\
$(3,4)_{-5/6}$  & $7/3$ & $XQ_{\left[i\right.}Q_{\left.j\right]}H$ & $X\bar DLHH$,$X\bar Q\bar LH$ \\
$(3,5)_{-1/3}$ & $7/3$ & $XQ_{\left[i\right.}Q_{\left.j\right]}H^\dagger H$ & $X\bar Q\bar LH^\dagger H$ \\ \hline\rule{0pt}{1.0em}
$(3,2)_{+13/6}$ & $8/3$ & & $X\bar ULH^\dagger H^\dagger$, $X\bar Q EH^\dagger H^\dagger$ \\
$(3,3)_{+5/3}$ & $8/3$ & $XD_{\left[i\right.}D_{\left.j\right]}H^\dagger H^\dagger$ &  $X\bar ULH^\dagger$, $X\bar QEH^\dagger$\\
$(3,4)_{+7/6}$ & $8/3$ & & $X\bar QEH^\dagger H$, $X\bar ULH^\dagger H$, $X\bar DLH^\dag H^\dag$ \\
$(3,5)_{+2/3}$ & $8/3$ & $XQ_{\left[i\right.}Q_{\left.j\right]}H^\dagger H^\dagger$ &  $X\bar Q\bar LH^\dagger H^\dagger$ \\ \hline\rule{0pt}{1.0em}
$(3,3)_{-7/3}$ & $10/3$ & $XU_{\left[i\right.}U_{\left.j\right]}HH$ & $X\bar D\bar EHH$ \\
$(3,5)_{-4/3}$ & $10/3$ & $XQ_{\left[i\right.}Q_{\left.j\right]}HH$  & $X\bar Q\bar LHH$\\
\hline\hline
\end{tabular}
\end{center}
\end{table}

\section{Direct searches for continuum pair production}\label{sec:direct}
Colored particles are pair-produced at the LHC via initial state gluons. In this section we study the direct searches for continuum pair production of color triplet $X_Q$.
The EM charge $Q$ dictates the possible decay modes and, subsequently, the experimental signatures. The $SU(2)_W$ quantum numbers are provisionally left out of the discussion.

Table~\ref{tab:scalardecay} summarizes the possible decay final states of  $X_Q$ for a given charge. We distinguish between two different decay topologies: 1) fully hadronic, in which $X_Q$ decays to two jets and possibly also $W$ bosons (we omit potential $jjh$ and $jjZ$ decay modes, as these are subdominant to an allowed $jj$ decay), and, 2) lepto-quark signature, in which $X_Q$ decays to a lepton (possibly a neutrino) and a jet.
\begin{table}[t]
\caption{ \it Possible Standard Model decay modes of $X_Q$ (or $\bar X_Q$), a scalar color (anti-)triplet with charge $Q$}
\label{tab:scalardecay}
\begin{center}
\begin{tabular}{ccc} \hline\hline
\rule{0pt}{1.0em}%
$Q$ & Hadronic & Lepto-quark  \\[2pt] \hline\hline
\rule{0pt}{1.0em}%
$1/3$ & $ud$, $uuW^-$ & $\bar ue^+$, $\bar d\nu$ \\
$2/3$ & $\bar d\bar d$ & $u\nu$, $de^+$ \\
$4/3$ & $uu$ & $\bar de^+$ \\
$5/3$ & $\bar d\bar dW^+$ & $u\nu W^+$, $de^+W^+$, $ue^+$ \\
$7/3$ & $uuW^+$ & $de^+W^+$ \\
$8/3$ & $\bar d\bar dW^+W^+$  & $ue^+W^+$ \\
$10/3$ & $uuW^+W^+$ & $de^+W^+W^+$ \\
\hline\hline
\end{tabular}
\end{center}
\end{table}
\begin{table}[h]
\caption{ \it Direct searches for $X_Q$ used in our analysis. The resulting bounds are shown in Fig.~\ref{fig:direct}. We use the notation $q$ for all six quark flavors, while $j=u,d,s,c$, and $\ell=e,\mu$.}
\label{tab:direct}
\begin{center}
\begin{tabular}{c|ccc} \hline\hline \rule{0pt}{1.0em}%
Diquark & $jj$ & $bj$ & $tj$, $tb$  \\[2pt] \hline \rule{0pt}{1.0em}%
Refs. & \cite{Khachatryan:2014lpa,Chatrchyan:2013izb,ATLAS:2012ds,Aad:2014aqa,ATLAS:084} & \cite{Aad:2016kww,ATLAS:2016yhq,Khachatryan:2014lpa} & \cite{Khachatryan:2016iqn} \cite{Chatrchyan:2013oba} \\
Comments  & RPV SUSY  & RPV SUSY & RPV SUSY search for $m\leq600$~GeV. \\
	         & searches     & searches     & $t^*\to tg$ search for higher masses.  \\
	         &                     &                    & We assume similar efficiencies for   \\
	         &                     &                    & high $p_T$ gluon and quark jets.  \\
\hline  \rule{0pt}{1.0em}%
 & $Wqq$ & $WWqq$ \\[2pt] \hline  \rule{0pt}{1.0em}%
Refs.  &\cite{Aaltonen:2011tq,Abazov:2011vy,Aad:2015tba}\cite{Aaltonen:2011vr,Aad:2015kqa,Chatrchyan:2012vu,CMS:2012ab}\cite{Aaltonen:2011tq,Aad:2015mba,ATLAS:2016sno,Chatrchyan:2012yea} &   \cite{Khachatryan:2016iqn,CMS:2016gvu,ATLAS:2016kjm} \\
Comments  & Recast $Wj,Wb$ and $Wt$ & Recast of \\
 & searches using $30\%-50\%$ efficiency & multilepton searches. \\
  &  reduction. See App.~\ref{app:3body}.  & See App.~\ref{app:4body}. \\
 \hline\hline \rule{0pt}{1.0em}%
Lepto-quark  & $\ell j$, $\ell b$ & $\ell t$ & $\tau j$ \\[2pt] \hline  \rule{0pt}{1.0em}%
Refs. &  \cite{Chatrchyan:2012vza,ATLAS:2012aq,Aad:2011ch,Aad:2015caa,Khachatryan:2015vaa,Aaboud:2016qeg,CMS:2016imw} &  \cite{CMS:2016gvu} &   \cite{Aad:2014wea,Aaboud:2016zdn} \\
Comments & 1st and 2nd generation & Recast of  & Recast of $\nu j$ \\
 &  lepto-quark searches & multilepton searches. & searches using $30\%-50\%$ \\
& w/o $b$ veto & See App.~\ref{app:ellt} & efficiency reduction. \\  \hline  \rule{0pt}{1.0em}%
                  & $\tau t$ &  $\tau b$ & \\ \hline  \rule{0pt}{1.0em}%
Refs. &  \cite{Khachatryan:2015bsa} &  \cite{Khachatryan:2014ura,CMS:2016hsa} &  \\
Comments &  3rd generation & 3rd generation & \\
		 & lepto-quark search & lepto-quark search & \\  \hline  \rule{0pt}{1.0em}%
                  & $\nu j$ &  $\nu t$ & $\nu b$ \\ \hline  \rule{0pt}{1.0em}%
Refs. &  \cite{Aad:2014wea,Aaboud:2016zdn} & \cite{Aad:2015caa,Aaboud:2016lwz} & \cite{Aad:2015caa,Aaboud:2016lwz}  \\
Comments & Standard SUSY searches & lepto-quark & lepto-quark \\
		 &  for squark pair & searches & searches \\
           	 &  with massless LSP & & \\
\hline\hline
\end{tabular}
\end{center}
\end{table}

Let us first analyze prompt signatures, highlighting the mass range $250{\rm \,GeV}\leq m_{X_Q}\leq1000$\,GeV.
For some $X_Q$ decay topologies, dedicated searches were carried out by ATLAS, CMS, or the Tevatron collaborations. These decay modes, along with the relevant searches, are summarized in Table~\ref{tab:direct}. However, some of the signatures we study have no dedicated experimental analysis.
We identify relevant searches which are sensitive to these topologies and estimate the corresponding efficiencies for our signal.
For this purpose we implement our model in FeynRules~\cite{Alloul:2013bka} and simulate the signal in MadGraph5~\cite{Alwall:2011uj} using Pythia 8~\cite{Sjostrand:2006za,Sjostrand:2007gs} for showering and hadroniztion. Detector effects are simulated in Delphes~\cite{deFavereau:2013fsa} using the standard configuration. We stress that, for the recasted channels, our results should be taken as an estimation only. A detailed description of our recast procedure can be found in Appendices~\ref{app:3body}, \ref{app:4body} and~\ref{app:ellt}.

\begin{figure}[h]
  \subfigure[ $qq$ final states]{%
    \includegraphics[width=0.45\textwidth]{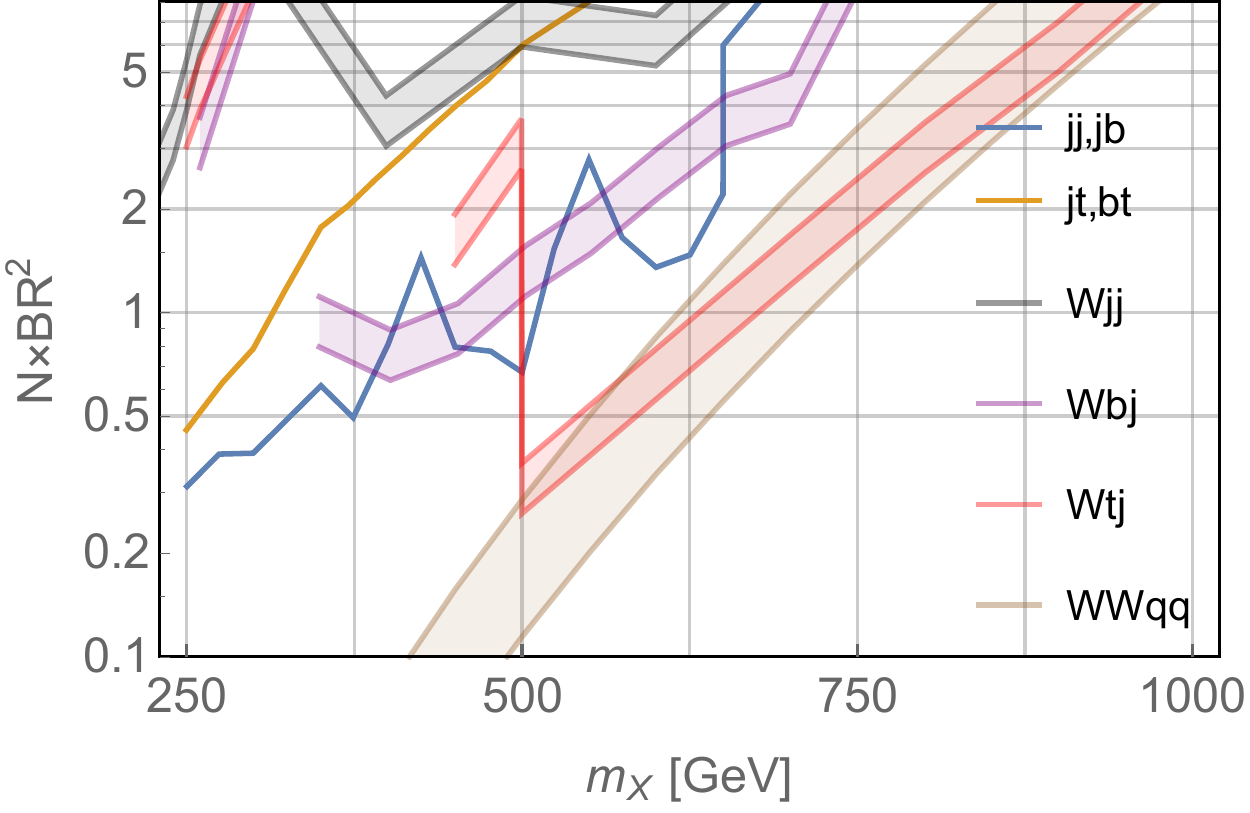}\label{fig:qq}
  }
  \quad
  \subfigure[ $q\ell$ final states]{%
    \includegraphics[width=0.45\textwidth]{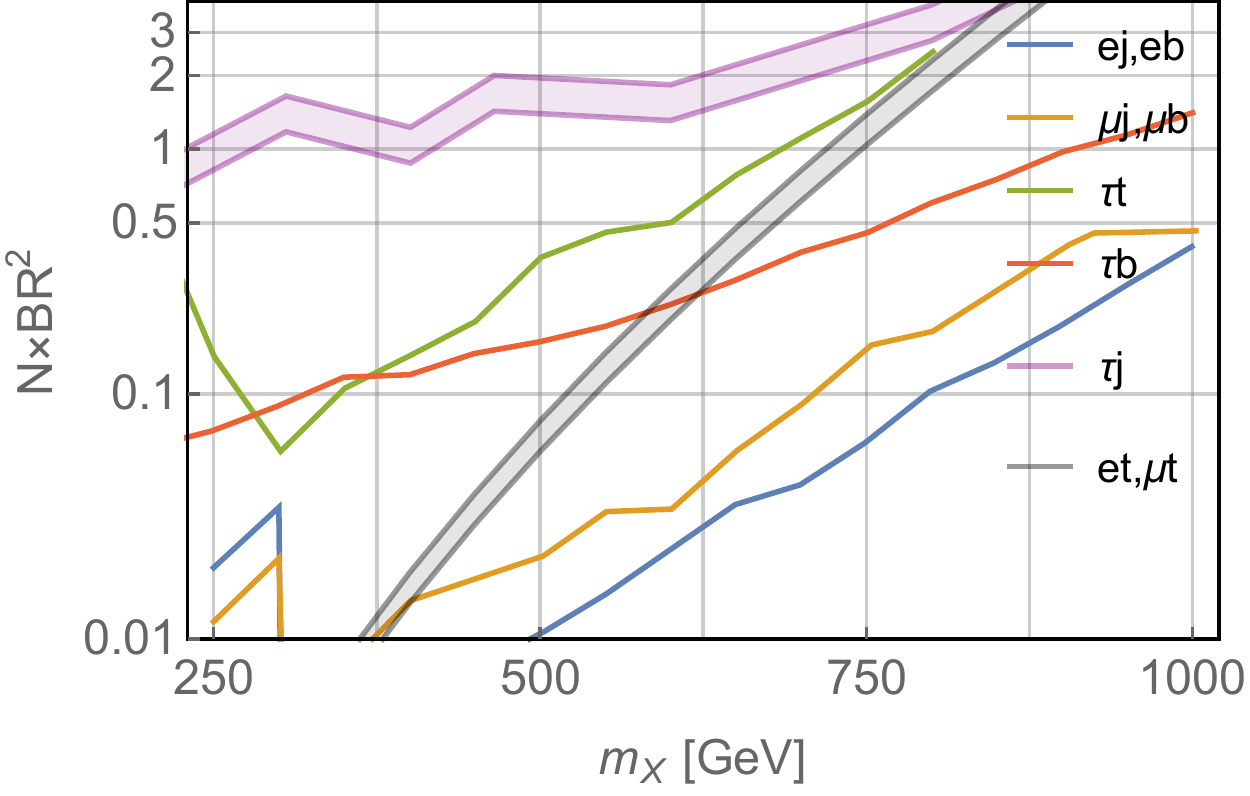}\label{fig:ql}
  }
    \quad
  \subfigure[ $q\nu$ final states]{%
    \includegraphics[width=0.45\textwidth]{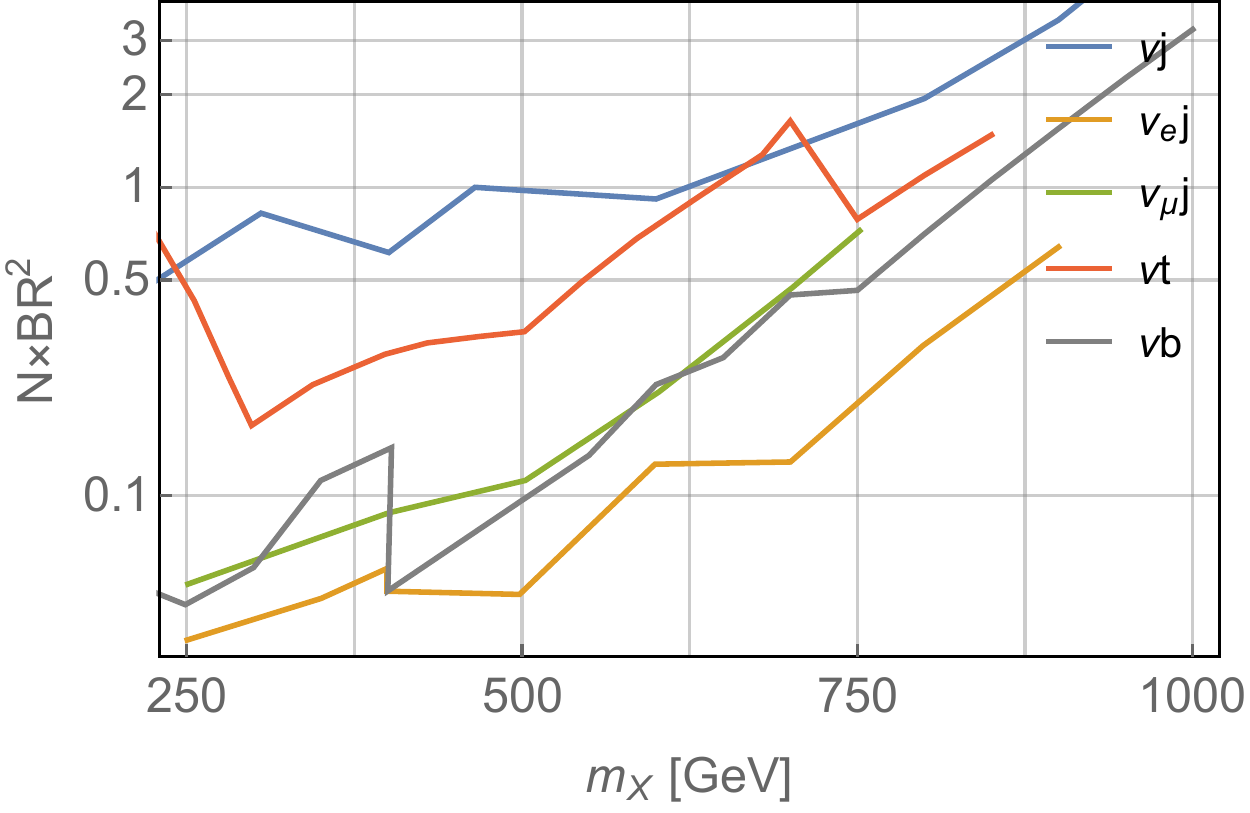}\label{fig:qnu}
  }
  \caption{ \label{fig:direct} \it Bounds from direct searches for $X_Q$ continuum pair production in the various final state topologies. The $y$ axes give $\sigma\times$BR$^2$ normalized to the pair production cross-section of a color-triplet scalar. Sharp features are caused by considering multiple searches for each channel; see the text for more details.}
\end{figure}

Our findings are presented in Fig.~\ref{fig:qq} for the dijet decays, Fig.~\ref{fig:ql} for the jet and charged lepton signals, and Fig.~\ref{fig:qnu} for the neutrino-jet topology. We also consider the case where a jet is replaced by heavy flavor quark. In each figure we show the current limit on the pair-production cross section times BR$^2$, normalized to the NLO+NLL cross section of a scalar colored triplet taken from~\cite{Beenakker:2010nq,Kramer:2012bx,Borschensky:2014cia}. Presented this way, when a single mode dominates the decay (namely ${\rm BR}=1$), the $y$ axis corresponds to the number of copies of the $X$ representation that are experimentally allowed.

An important ingredient for collider phenomenology is the lifetime of $X_Q$. Non-prompt decays are studied by the experimental collaborations in dedicated searches, leading to bounds in the ballpark of $m_{X_Q}\gtrsim700-900$~GeV for color-triplet scalars.
Refs.~\cite{Liu:2015bma,Csaki:2015uza} analyzed displaced signatures in the context of RPV SUSY models. They find that $X_Q$ in the mass range of $100-1000$\,GeV, decaying to dijet, or to a jet and a charged lepton, or to a jet and a neutrino, would not be captured by the displaced-track searches if its mean-free path is less than $0.3-10$~mm. While the exact number depends on the particle mass and decay mode, we conservatively use in the following $0.3$~mm as an upper bound on a two-body decay length. We are not aware of any dedicated analysis for displaced signature of a three- (or four-) body final state. We estimate that the larger multiplicity of the final objects would increase the efficiency of these searches at high $m_{X_Q}$, while the low $m_{X_Q}$ regime will suffer from the typically lower energy carried  by each final object. Over all, we expect that the sensitivity is comparable to the other topologies, and so we consider $c\tau\lesssim1$~mm for three-body decay. We then apply the following 'promptness` requirements on $X_Q$ decay rates:
\beqa\label{eq:prompt}
&&\Gamma_{\rm 2-body}\gtrsim7\times10^{-13}{\rm \,GeV}\,,\;\;\;\Gamma_{\rm 3-body}\gtrsim2\times10^{-13}{\rm \,GeV}\,,
\eeqa
which translate into lower bounds on the Yukawa coupling of $X_Q$ to SM states.

Concluding this section, we learn the following:
\begin{itemize}
\item The lepto-quark topology is strongly constrained by direct searches. As can be seen in Fig.~\ref{fig:ql}, none of the decay modes in this category allows for more than two states below $m_{X_Q}\simeq750$~GeV.
\item The neutrino-quark topology is subject to the standard SUSY searches for jet and missing energy. As can be seen in Fig.~\ref{fig:qnu}, the corresponding bounds on $m_{X_Q}$ are even stronger than in the $jl$ category.
\item The hadronic decay modes are significantly less constrained by direct searches. This is expected given the large QCD backgrounds at the LHC. 
\item A $Wjj$ signature is poorly constrained by the LHC. As we show below, such topology could be the signature of multiple states which undergo cascade $SU(2)_W$ decays. This is an important gap in the LHC coverage for colored new particles which motivates dedicated searches for this decay topology.
\end{itemize}

\section{Mass splitting and cascade decays}\label{sec:cascade}
In general, two members of an $SU(2)_W$ multiplet with EM charges $Q$ and $Q'$ are split in mass. Tree level mass splittings are induced by the $\lambda^\prime_{XH}$ term:
\be(m_Q^2-m^2_{Q^\prime})^{\rm tree}&=&-\frac{\lambda^\prime_{XH}v^2}{4}(Q-Q^\prime)\no\\
\Rightarrow(m_Q-m_{Q^\prime})^{\rm tree}
&\simeq&1.7\ {\rm GeV}\times(Q^\prime-Q)\left(\frac{\lambda^\prime_{XH}}{0.1}\right)\left(\frac{m_X}{450~{\rm GeV}}\right)^{-1}.\ee
Mass splittings also arise through electroweak gauge boson loops from the kinetic term $(D_\mu X)^\dagger(D^\mu X)$~\cite{Cirelli:2005uq}:
\beqa
(m_Q-m_{Q^\prime})^{1-{\rm loop}}&=&\frac{\alpha m_Z}{2}
\left\{(Q^2-Q^{\prime2})\tilde f(x_Z)+(Q-Q^\prime)(Q+Q^\prime-2Y)\frac{1}{s_W^2}\left[c_W\tilde f(x_W)-\tilde f(x_Z)\right]\right\}\no\\
\Rightarrow(m_Q-m_{Q^\prime})^{1-{\rm loop}}&\simeq&-0.15~{\rm GeV}\times(Q^\prime-Q)\left(Q^{\prime}+Q+2.3Y\right),
\eeqa
where $s_W^2\equiv\sin^2\theta_W$, $c_W\equiv\cos\theta_W$, $x_V=m_V/m_X$, and $\tilde f(x)=-\frac{1}{8\pi}(2x^3\log x+(x^2-4)^{3/2}\log[(x^2-2-x\sqrt{x^2-4})/2])=1-\frac{x}{3}+\mathcal{O}(x^2)$.

Assuming no fine-tuned cancelations between the tree and loop contributions, a mass splitting of at least ${\cal O}(100\ {\rm MeV})$ between adjacent members of the multiplet ($Q=Q^\prime+1$) is unavoidable. Much larger splittings are possible, depending on the value of $\lambda^\prime_{XH}$.
If the tree contribution dominates, the splitting can be of either sign, and the lightest colored scalar is the one with either the highest or the lowest $Q$.

The mass splitting between the members of an $SU(2)_W$ multiplet leads to $W$-mediated decays within the multiplet, $X_m\to X_{m\pm1}W^{\mp(*)}$. (Note that we change notations in this section from $X_Q$ to $X_m$, with $Q=m+Y$.) For the three-body decay,  $X_m\to X_{m+1}f\overline{f^\prime}$, with massless fermions, we obtain
\beqa\label{eq:xwll}
\Gamma(X_m\to X_{m+1}f\overline{f^\prime})&&=\frac{G_F^2}{15\pi^3}\left(j-m\right)\left(j+m+1\right)(\Delta M)^5\nonumber\\
&&\simeq3\times10^{-13}\ {\rm GeV}\left(\frac{\Delta M}{1\ {\rm GeV}}\right)^5\left(j-m\right)\left(j+m+1\right)\,.
\eeqa
If $\Delta M>m_\pi$, we have the two body decay $X_m\to X_{m+1}\pi^-$, in which case
\beqa\label{eq:xwpi}
\Gamma(X_m\to X_{m+1}\pi^-)&&=\frac{G_F^2}{\pi}\left(j-m\right)\left(j+m+1\right)(\Delta M)^3 f_\pi^2\sqrt{1-\frac{m_\pi^2}{(\Delta M)^2}}\nonumber\\
&&\simeq7\times10^{-13}\ {\rm GeV}\left(\frac{\Delta M}{1\ {\rm GeV}}\right)^3\left(j-m\right)\left(j+m+1\right)\,.
\eeqa
For $m=-1$ we recover the results of Ref.~\cite{Cirelli:2005uq}. We do not consider $\Delta M>m_W$.

To determine the phenomenological significance of these decays (for all but the lightest member of the multiplet), we need to compare their rate to those of the Yukawa mediated decays. We will do so in the next section.

\section{A model example: $X(3,4)_{+\frac{1}{6}}$}\label{sec:cases}
In the following we discuss the model example $X\sim(3,4)_{+{1}/{6}}$, containing a state with $Q=+5/3$ as the highest charge state. We assign $X$ zero lepton number which, given our assumptions in Sec.~\ref{sec:model}, forces $X_{+5/3}$ to decay into the hadronic three body state $\bar d_i\bar d_j W^+$ via the operator
\beq\label{eq:Yquartet}
\mathcal{L}_{Y_X}=\frac{Y^{QQ}_{ij}}{2}Q_{\left[i\right.} Q_{\left.j\right]}H^\dagger X+{\rm h.c.}\,,
\eeq
with $Y^{QQ}_{ij}$ antisymmetric in the flavor indices $i,j$, and of dimension mass$^{-1}$.

We consider two specific scenarios:
\begin{itemize}
\item Case A: degenerate $X_Q$ states.
\item Case B: non-degenerate  $X_Q$ states.
\end{itemize}
We now show that these two cases exhibit distinct phenomenology.

\subsection{Degenerate $SU(2)_W$-quartet}
The Lagrangian (\ref{eq:Yquartet}) gives the following component interaction terms (to leading order in CKM rotation) for the four multiplet members (with $Q=+5/3,+2/3,-1/3,-4/3$):
\beqa
\mathcal{L}&&=Y^{QQ}_{ij}\left[X_{+5/3}d_id_jh^-+X_{-4/3}\left(\frac{v}{\sqrt{2}}u_iu_j+\frac{1}{\sqrt{2}}u_iu_j(h+i\rho)\right)\right.\no\\
&&\left.+X_{+2/3}\left(\frac{v}{\sqrt{6}}d_id_j+\frac{1}{\sqrt{6}}d_id_j(h+i\rho)
 -\frac{1}{\sqrt{3}}d_iu_jh^--\frac{1}{\sqrt{3}}u_id_jh^-\right)\right.\no\\
&&\left.+X_{-1/3}\left(\frac{v}{\sqrt{6}}d_iu_j+\frac{v}{\sqrt{6}}u_id_j
 +\frac{1}{\sqrt{6}}d_iu_jh+\frac{1}{\sqrt{6}}u_id_jh-\frac{1}{\sqrt{3}}u_iu_jh^-\right)\right]+{\rm h.c.}\,.
\eeqa
(We work in the Feynman-'t Hooft gauge $\xi=1$ as to straightforwardly keep track of longitudinal $W^+,Z$ modes.) These terms allow two body decays of $X_{+2/3}$, $X_{-1/3}$ and $X_{-4/3}$:
\beq\label{eq:jj}
\Gamma\left(X_Q\rightarrow\bar q_{Li}\bar q_{Lj}\right)
=c_Q\,\abs{Y^{QQ}_{ij}}^2\frac{3v^2}{16\pi}m_X
\lambda\left[m_i^2,m_j^2,m_X^2\right]^{1/2}\beta\left[m_i+m_j,m_X\right],
\eeq
where $c_Q=\frac{1}{3},\frac{1}{3},1$ for $Q=+2/3,-1/3,-4/3$, respectively, $\lambda[x,y,z]=\left(1-x/z-y/z\right)^2-4xy/z^2$, and $\beta[x,y]=\left(1-x^2/y^2\right)$.
They also allow three body decays of all four members:
\beq\label{eq:wqq}
\Gamma\left(X\rightarrow\phi\bar q_{Li}\bar q_{Lj}\right)=\tilde c_Q\,
\abs{Y^{QQ}_{ij}}^2\frac{m_X^2}{512\pi^3}m_X
\left[1+\frac{m_\phi^2}{m_X^2}\left(9+6\log\frac{m_\phi^2}{m_X^2}\right)
-6\frac{m_{i}^2}{m_X^2}+\mathcal{O}\left(\frac{m_{i,\phi}}{m_X}\right)^4\right]\,,
\eeq
Here $\tilde c_Q=1,\frac{1}{3},\frac{1}{6},\frac{1}{2}$ for $Q=+5/3,+2/3,-1/3,-4/3$, respectively, and we take $m_j=0$. The boson $\phi$ is a neutral Higgs or a longitudinal gauge boson.

For $m_X\lesssim8$~TeV, the two-body decays of Eq.~(\ref{eq:jj}), where available, dominate over the three-body decays of Eq.~(\ref{eq:wqq}). If the $Y^{QQ}$ term dominates the decay rates of all members of the quartet, then
\beqa
&&{\rm BR}[X_{+5/3}\to \bar d_i\bar d_j W^+]\simeq1\,,\;\;\;{\rm BR}[X_{-4/3}\to \bar u_i\bar u_j]\simeq1\,,\;\;\;{\rm BR}[X_{+2/3}\to \bar d_i\bar d_j]\simeq1,\no\\
&&{\rm BR}[X_{-1/3}\to \bar u_i\bar d_j]={\rm BR}[X_{-1/3}\to \bar d_i\bar u_j]\simeq\frac{1}{2}\,.
\eeqa
For $i,j=1,2$, we have three states decaying into a $jj$ final state, and one state decaying into a $Wjj$ topology. This is allowed for $m_X\gtrsim630$\,GeV. For $i=3$, we have effectively $1.5$ members decaying into $jb$ and $jt$ each. Looking at $N\times$BR$^2=1.25$ in Figure~\ref{fig:qq} we conclude that $m_{X_Q}=520$~GeV is a viable possibility. We use this mass as our benchmark point in the following. To guarantee prompt decay of $X_{+5/3}$ we impose
\beqa\label{eq:yukminII}
Y^{QQ}_{3j}\gtrsim7\times10^{-9}\,{\rm GeV}^{-1}(520{\rm \,GeV}/m_X)^{3/2}\,.
\eeqa
%

\subsection{Non-degenerate $SU(2)_W$-quartet}

Mass splitting between the members of the quartet allow for fast cascade decays of the three heavier ones.
In order to establish their phenomenological relevance one needs to compare the rate of these weak decays with the rate of the Yukawa mediated decay modes, which depend on the dimensional coupling  $Y^{QQ}_{ij}$ , Eqs.~\eqref{eq:jj} and~\eqref{eq:wqq}. The dominant terms need to induce prompt decays for all the members of the $X$ multiplet.  We distinguish between two cases:
\begin{enumerate}
\item $X_{-4/3}$ is the lightest: In this case, either all states decay dominantly via their Yukawa coupling, or the $Q=+5/3$ state (and possibly also the $Q=+2/3$ and $Q=-1/3$ states) decay via $W$-mediated cascade decays. In either case, we have at least three color-triplet states decaying into two jets. The mass of the lightest state should then be similar to the mass considered in the degenerate quartet scenario.
\item $X_{+5/3}$ is the lightest: In this case, $X_{+5/3}$ decays to a $\bar d_i\bar d_jW^+$ final state. As concerns the three heavier states, they can either decay into two jets, or cascade into the $X_{+5/3}$ state. The latter would lead to effectively four states decaying to $Wqq$ in the final state, assuming the other cascade products are too soft to be detected (this is the case for a few GeV splitting). As far as direct searches for continuum pair production are concerned, we estimate the sensitivity of top-partner searches at the Tevatron~\cite{Aaltonen:2011tq} and find that, in this case, $X_{+5/3}$ can be as light as $250$~GeV. As we will see next, the direct searches for an $X_QX_Q^\dagger$ bound state place a stronger limit, of $m_X\gtrsim450$~GeV, with a corresponding lower bound on the Yukawa coupling,
\beqa\label{eq:yukminIII}
Y^{QQ}_{\rm min}\simeq9.1\times10^{-9}\,{\rm GeV}^{-1}(450{\rm \,GeV}/m_X)^{3/2},
\eeqa
to ensure its prompt decay. Using $Y^{QQ}_{\rm min}$ as a convenient reference, and recalling that the two-body decay rate is faster than the three-body one for $m_X\lesssim8$~TeV,
a mass splitting of
\beqa
\Delta M\gtrsim2.7{\rm \,GeV}\left(\frac{450{\rm ~GeV}}{m_X}\right)^{2/5}\left(\frac{Y^{QQ}}{Y^{QQ}_{\rm min}}\right)^{2/5}
\eeqa
between two `adjacent' members of the multiplet would effectively cause the four members of $X(3,4)_{+1/6}$ decay to $Wqq$ final states. The precise coefficient varies a little between the different $SU(2)_W$ members.
\end{enumerate}

We therefore consider, for our second scenario, the following spectrum:
\beqa
m_{X_{+5/3}}=450~{\rm GeV},\;\;\;m_{X_{+2/3}}=452.8~{\rm GeV},\;\;\;m_{X_{-1/3}}=455.7~{\rm GeV},\;\;\;m_{X_{-4/3}}=458.5~{\rm GeV}\,,
\eeqa
which is the result of $\lambda_{XH}^\prime=0.17$.

\section{$SU(2)_W$ phenomenology}\label{sec:SU2}
In this section we explore the distinct phenomenology of colored $SU(2)_W$ non-singlet scalars.

\subsection{Electroweak precision measurements (EWPM)}
Large mass splitting within an $SU(2)_W$ multiplet is constrained by EWPM. Specifically, it modifies the oblique $T$ and $S$ parameters~\cite{Peskin:1991sw}, where the leading effect comes from generating the dimension six operators $\mathcal{O}_T$ and $\mathcal{O}_{WB}$ (see App.~\ref{app:EFTX} for the definition of these operators). For an $(R,n)_Y$ representation, we have
\beqa
T&&=\frac{v^2}{\alpha}c_T=\left(\frac{v^2}{4608\pi^2\alpha}\right)\left(\frac{\lambda_{XH}^{\prime 2}}{m_X^2}\right)Rn(n^2-1)
=4.4\times10^{-3}\left(\frac{\lambda^{\prime}_{XH}}{0.17}\right)^2\times\left(\frac{Rn(n^2-1)}{180}\right)\left(\frac{450\,{\rm GeV}}{m_X}\right)^2,\no\\
S&&=16\pi v^2c_{WB}=\left(\frac{v^2}{144\pi}\right)\left(\frac{\lambda_{XH}^\prime}{m_X^2}\right)Rn(n^2-1)Y
=3.4\times10^{-3}\left(\frac{\lambda^{\prime}_{XH}}{0.17}\right)\times\left(\frac{Rn(n^2-1)Y}{30}\right)\left(\frac{450\,{\rm GeV}}{m_X}\right)^2\,,
\eeqa
where in the second equation of each line we normalize to the quantum numbers of $X(3,4)_{+1/6}$ and to the value of $\lambda^\prime_{XH}$ which we use for case B in Sec.~\ref{sec:cases}. The EWPM constraints read (for $U=0$) \cite{Baak:2014ora}
\beq
T=0.10\pm0.07,\ \ \ S=0.06\pm0.09\,,
\eeq
with correlation of $\rho=0.91$. Using one dimensional $\chi^2(\lambda^\prime)$ function we find that $\abs{m_{X_Q}-m_{X_{Q\pm1}}}\lesssim13-16$\,GeV is allowed around $450$~GeV, where a positive (negative) $\lambda^\prime_{XH}$ implies that $X_{+5/3}$ ($X_{-4/3}$) is the lightest member of the multiplet. Clearly, EWPM allow the mass splitting we consider in case B.

In the limit of custodial symmetry, modifications to the EW vacuum polarization amplitude alter the oblique $Y$ and $W$ parameters~\cite{Barbieri:2004qk}. These are primarily encoded in the dimension six operators $\mathcal{O}_{2B},\mathcal{O}_{2W}$:
\beqa
Y=&&2m_W^2c_{2B}=\frac{g^{\prime2}}{240\pi^2}\frac{m_W^2}{m_X^2}RnY^2\simeq5.7\times10^{-7}\times\left(\frac{RnY^2}{1/3}\right)\left(\frac{450{\rm \,GeV}}{m_X}\right)^2\,,\no\\
W=&&2m_W^2c_{2W}=\frac{g^2}{2880\pi^2}\frac{m_W^2}{m_X^2}Rn\left(n^2-1\right)\simeq8.9\times10^{-5}\times\left(\frac{Rn(n^2-1)}{180}\right)\left(\frac{450\,{\rm GeV}}{m_X}\right)^2\,.
\eeqa
These contributions to $Y$ and $W$ are below the current sensitivity of LEP (see, {\it e.g.}, table 4 of~\cite{Alves:2014cda}) and the LHC~\cite{Farina:2016rws}. The values we take for the various coupling constants are listed in App.~\ref{app:RGE}.

\subsection{Gauge coupling running}

The presence of $X\sim(R,n)_Y$ modifies the running of the gauge coupling constants. We describe this effect, at one-loop level, in App.~\ref{app:RGE}. In particular, high $SU(2)_W$ representations change significantly the running of $\alpha_2$. For instance, color-triplet in the quartet (or higher) representation of $SU(2)_W$ flips the sign of the $SU(2)_W$ beta function. In particular, for $X(3,5)_Y$, $\alpha_2$ becomes non-perturbative at $\mu\simeq10^{15}$ GeV. Since the decay of $X$ already requires some cut-off at a lower scale, this is insignificant to our study.

Additional probe for the running of EW gauge coupling is the differential distribution of Drell-Yan processes at various energies, as was proposed in Ref.~\cite{Alves:2014cda}. Ref.~\cite{Goertz:2016iwa} finds that for $m_\psi=520$\,GeV, $N_\psi Q^2\geq46$ is excluded at the $2\sigma$ level, where $N_\psi$ is the number of copies of vector-like fermions transforming as $\psi\sim(3,1)_Q$. This scenario would generate a $23\%$ ($50\%$) relative increase in the Drell-Yan rate at $m_{\ell\ell}=1\,(1.5)$~TeV, which excludes $b_2^X\leq -46$. In our model example of Sec.~\ref{sec:cases}, $b_2^X=-Rn(n^2-1)/36=-5$, clearly within bounds. A more recent analysis done in Ref.~\cite{Farina:2016rws} yields the same conclusion.

\subsection{Additional constraints}

{\bf SM Higgs couplings:}
Integrating out $X$ generates dimension six effective operators involving the Higgs field. These, in turn, modify the Higgs couplings to fermions and gauge-bosons with respect to their SM values. LHC Higgs data constrain these modifications, resulting in bounds on the quartic couplings $\lambda_{XH}$ and $\lambda^\prime_{XH}$. At present, EWPM induce stronger constraints on $\lambda^\prime_{XH}$. The Higgs data do constrain $\lambda_{XH}$, but this coupling is not directly relevant to our analysis. We present our numerical results of the Higgs data for $X(3,4)_{+1/6}$ in App.~\ref{app:Higgs}, and the resulting minor effects on the various $S\to VV$ decays in App.~\ref{app:rates}.
%

{\bf Scalar quartic coupling running:}
In addition to modifying the SM Higgs couplings to fermions and gauge bosons, the presence of $X$ changes the running of the SM Higgs couplings. We calculate these effects in App.~\ref{app:RGEquartic}. We find that no dangerous runaway behavior is generated.
The same conclusion holds for the $X$ quartic coupling, and the mixed couplings $\lambda_{XH}$ and $\lambda_{XH}^\prime$.

\section{QCD bound state}\label{sec:bound}

In the previous sections we obtained constraints from both direct and indirect probes on the existence of exotic colored scalars. The interesting result is that these constraints
can be quite mild, allowing rather light colored scalars. For example, as demonstrated by the non-degenerate quartet scenario (case B in Sec.~\ref{sec:cases}), the data still allow four colored states with $m_X\simeq250$~GeV. In this section we study another way to discover light colored scalars, which might go first through the observation of their QCD bound state~\cite{Kats:2012ym,Kats:2016kuz}. Moreover, constraints derived from bound state searches are less model dependent, in the sense that they do not depend on the decay mode of $X$.

A pair of $X_QX_Q^\dagger$ near threshold can form a QCD bound state, which we denote by $S_Q$. If the decay rates of its constituents are smaller than $\Gamma_{S_Q}$, and its width is smaller than the respective binding energy, $S_Q$ can be seen as a resonance as it annihilates into SM particles. For a review we refer the reader to Ref.~\cite{Kahawala:2011pc} and references therein. Heavy constituents exhibit Coulomb-like potential with a binding energy
\beqa
E_b=-\frac{1}{4n_E^2}[C_2(R)]^2\overline\alpha_s^2m_X\,,
\eeqa
where $n_E$ is the excitation index ($n=1$ is the ground state), $\overline\alpha_s$ is the strong coupling evaluated at the bound-state typical scale (for which we use, following Ref.~\cite{Kahawala:2011pc}, the Bohr radius) and $C_2(R)$ is the quadratic $SU(3)$ Casimir of representation $R$, with $C_2(3)=4/3$. We assume that the resulting bound state is an $SU(3)_C$ singlet. The mass of $S$ is $m_S=2m_X+E_b$.

The condition that pair annihilation dominates the decay of $S_Q$ reads
\be\label{eq:boucon}
2\Gamma_X < \Gamma_{S_X}= 2 \times 10^{-5}  m_X.
\ee
The RHS is well above the lower bounds in Eq. (\ref{eq:prompt}). In fact, (\ref{eq:boucon}) is fulfilled quite generally by the exotic states on which we focus the analysis. The argument goes as follows. Suppose that $X$ decays into a two fermion final state, with effective coupling $y$. The condition (\ref{eq:boucon}) translates into $y<10^{-2}$. If the effective coupling comes from a dimension $d$ operator, we have $y=\hat y(v/\Lambda)^{d-4}$, where $\hat y$ is dimensionless and $\Lambda$ is the scale of new physics. We assume perturbativity ($\hat y\lesssim1$), and a NP scale that is not very low ($\Lambda\gtrsim10$ TeV). Then, for $d=6$ operators, the condition is always fulfilled. For $d=5$ operators it is not fulfilled only in a small region of parameter space where $\Lambda\lesssim25$ TeV and $\hat y>0.4$. Fully hadronic decays via renormalizable operators ($d=4$) are possible only in a single case of $SU(2)_W$ non-singlet, that is $X(3,3)_{-1/3}$, and even then the condition is fulfilled for $\hat y<0.01$. The condition (\ref{eq:boucon}) applies in all cases of dominant three body final state. We conclude that the search for bound states is truly a generic tool to look for exotic colored scalars~\cite{Kats:2012ym}. 

The quantum numbers of $X$ determine the gluon fusion (ggF) production cross section of $S$ as well as its decay rates into pairs of vector bosons: $gg$, $\gamma\gamma$, $ZZ$, $Z\gamma$ and $WW$. Assuming that the $X+X^\dagger$ production is dominated by ggF, and that there are no additional decay modes that give a significant contribution to the total width of $S$, then $\sigma(pp\to S)\times{\rm BR}(S\to V_1V_2)$ is predicted. The ggF partonic production cross section is given by
\beqa
\hat\sigma_{gg\to S}&&=\frac{\pi^2}{8}\frac{\Gamma(S\to gg)}{m_S}\delta(\hat s-m_S^2)\,.
\eeqa
We convolute $\hat\sigma$ with the partonic luminosity function
\beqa
\sigma=\frac{\hat\sigma}{m_S^2} \frac{\tau{\rm d}\mathcal{L}}{{\rm d}\tau}
\eeqa
evaluated at $\tau=m_S^2/s$, where $\sqrt{s}$ is the CoM energy. For the various two-body decay rates, we use (see~\cite{Kats:2012ym} and references therein)
\beq
\Gamma(S_Q\to V_1V_2)=\frac{R}{8\pi(1+\delta_{V_1V_2})}\frac{|\psi(0)|^2}{m_S^2}|{\overline{\cal M}}_{V_1V_2}|^2\lambda^{1/2}(m_S^2,m_{V_1}^2,m_{V_2}^2),
\eeq
where $\lambda[x,y,z]$ is defined below Eq.~\eqref{eq:jj}, and $\psi(0)$ is the joint wave function of $X_QX_Q^\dagger$ at the origin, which controls the probability to form a bound state, and is given by
\beqa
\abs{\psi(0}^2=\frac{[C_2(R)]^3\overline\alpha_s^3m_X^3}{8\pi n^3}\,.
\eeqa

The full expressions for $|{\overline{\cal M}}_{V_1V_2}|^2$ can be found in App.~\ref{app:rates}. We provide here the ratios between the different decay rates of $S_Q$ (with $Q=m+Y$), denoting $R^Q_{X/Y}=\Gamma(S_Q\to X)/\Gamma(S_Q\to Y)$, and neglecting contributions proportional to $\tilde\lambda^m_{XH}=\lambda_{XH}-(m/2)\lambda^\prime_{XH}$ and phase space suppressions:
\beqa
R^Q_{gg/\gamma\gamma}&&=\frac{C_2(R)^2\alpha_s^2}{8Q^4\alpha_{\rm EM}^2}\,,\;\;\;\;\;\;\;\;\;\;\;\;
R^Q_{ZZ/\gamma\gamma}=\frac{[m-Qs_W^2]^4}{s_W^4 c_W^4 Q^4}\,,\nonumber\\
R^Q_{Z\gamma/\gamma\gamma}&&=\frac{2[m-Qs_W^2]^2}{s_W^2 c_W^2Q^2}\,,\;\;\;\;\;\;\;
R^Q_{WW/\gamma\gamma}=\frac{(n^2-1-4m^2)^2}{32s_W^4Q^4}\,.
\eeqa
In the limit of small mass splitting, the various $V_1V_2$ signals depend on the sum of the branching ratio of each member, rather than on the sum of $R_Q$.
They are the same if the total width of all the $S_Q$ members is equal, which is the case if the digluon mode dominates the total width.
In Tab.~\ref{tab:VVratios} we calculate the ratios between the different $V_1V_2$ signals, summing over all $S_Q$'s. Note that the running of the gauge coupling slightly modifies the numerical values of these ratios for various bound state masses. For concreteness, we quote these values at $m_S=800$~GeV, and denote
\beqa
R_{X/Y}=\frac{\sum_Q{\rm Br}[S_Q\to X]}{\sum_Q{\rm Br}[S_Q\to Y]}\,.
\eeqa
We further specify, in Tab.~\ref{tab:VVratios}, $\sigma_{\gamma\gamma}^{13}$, the expected diphoton signal at the 13~TeV LHC for the various representations we consider, taking $m_S=800$~GeV.
\begin{table}[b]
\caption{ \it The $\sigma^{13}_{\gamma\gamma}$ cross section for $m_S=800$~GeV and the ratios between $S\to V_1V_2$ and the diphoton signals for the various $SU(3)_C\times SU(2)_W\times U(1)_Y$ representations. The singlet values are valid for small hypercharge, assuming the digluon decay mode dominates the total width of $S$.}
\label{tab:VVratios}
\begin{center}
\begin{tabular}{lcccccc} \hline\hline
\rule{0pt}{1.0em}%
$(R,n)_Y$ & $\abs{Q}_{\rm high}$ & $\sigma^{13}_{\gamma\gamma}$ [fb] &  $R_{WW/\gamma\gamma}$ &
$R_{ZZ/\gamma\gamma}$ & $R_{Z\gamma/\gamma\gamma}$ & $R_{gg/\gamma\gamma}$ \\[2pt] \hline\hline
\rule{0pt}{1.0em}%
$(3,1)_Y$   & $|Y|$ & $0.48 Y^4$ &  0 & 0.09 & 0.6 & $30Y^{-4}$ \\ \hline\rule{0pt}{1.0em}%
$(3,2)_{+1/6}$ & 2/3 & $0.09$ & $22$& $6.8$& $3.8$ & $286$ \\ \hline\rule{0pt}{1.0em}%
$(3,2)_{-5/6}$ & 4/3 & $1.3$ & $1.4$ & $1.1$ & $0.3$ & 19 \\
$(3,3)_{-1/3}$ & 4/3 & $0.9$ & $15$ & $6.8$ & $3.9$ & 26 \\
 \hline\rule{0pt}{1.0em}%
$(3,2)_{+7/6}$ & 5/3 & $3.0$ & $0.6$ & $0.7$ & $0.3$ & 7.6 \\
$(3,3)_{+2/3}$ & 5/3& $1.9$ & $6.7$ & $3.4$ & $1.6$ & 11 \\
$(3,4)_{+1/6}$ & 5/3& $1.1$ & $25$ & $10.7$ & $6.1$ & 11\\
 \hline\rule{0pt}{1.0em}%
$(3,2)_{-11/6}$ & 7/3 & $8.0$ & $0.1$ & $0.4$ & $0.4$ & 1.8 \\
$(3,3)_{-4/3}$ & 7/3& $6.1$ & $1.8$ & $1.3$ & $0.5$ & 2.8\\
$(3,4)_{-5/6}$ & 7/3 & $3.1$ & $8.8$ & $4.0$ & $2.0$ & 3.8 \\
$(3,5)_{-1/3}$ & 7/3& $1.7$ & $25$ & $9.7$ & $5.6$ & 4.0 \\
\hline\rule{0pt}{1.0em}%
$(3,2)_{+13/6}$ & 8/3& $10$ & $0.08$ & $0.3$ & $0.4$ & 1.0 \\
$(3,3)_{+5/3}$ & 8/3& $9.1$ & $1.1$ & $0.9$ & $0.4$ & 1.6 \\
$(3,4)_{+7/6}$ & 8/3& $5.0$ & $5.2$ & $2.5$ & $1.1$ & 2.2 \\
$(3,5)_{+2/3}$ & 8/3& $2.5$ & $17$ & $6.6$ & $3.6$ & 2.7 \\
 \hline\rule{0pt}{1.0em}%
$(3,3)_{-7/3}$ & 10/3& $15$ & $0.4$ & $0.6$ & $0.4$ & 0.6 \\
$(3,5)_{-4/3}$ & 10/3& $6.0$ & $7.0$ & $3.0$ & $1.4$ & 1.1 \\
\hline\hline
\end{tabular}
\end{center}
\end{table}
Bound state composed of $SU(2)_W$ non-singlet exhibit several interesting features, which we discuss next.

\subsection{Diphoton signature}
\begin{figure}[h!]
  \subfigure[\ $SU(2)_W$ singlet]{%
    \includegraphics[width=0.4\textwidth]{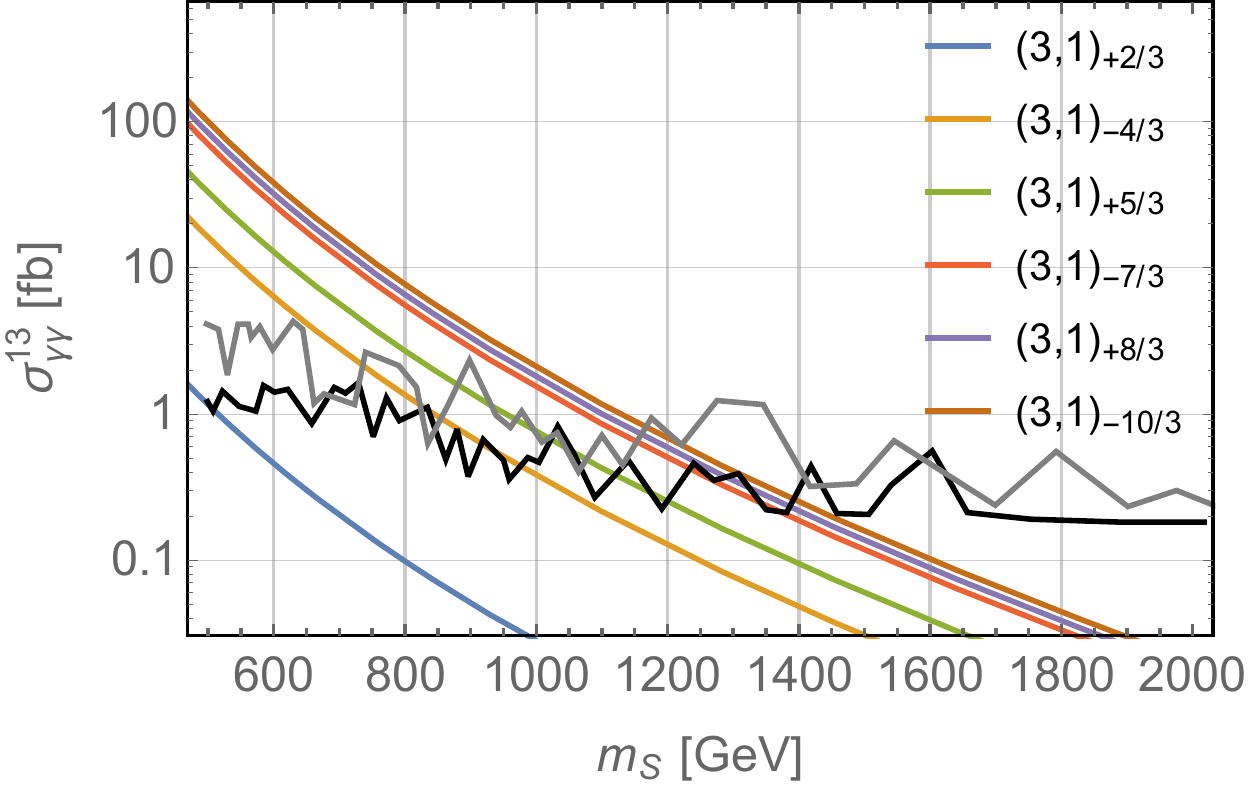}\label{fig:sing}
  }
  \quad
  \subfigure[\ $SU(2)_W$ high rep.]{%
    \includegraphics[width=0.4\textwidth]{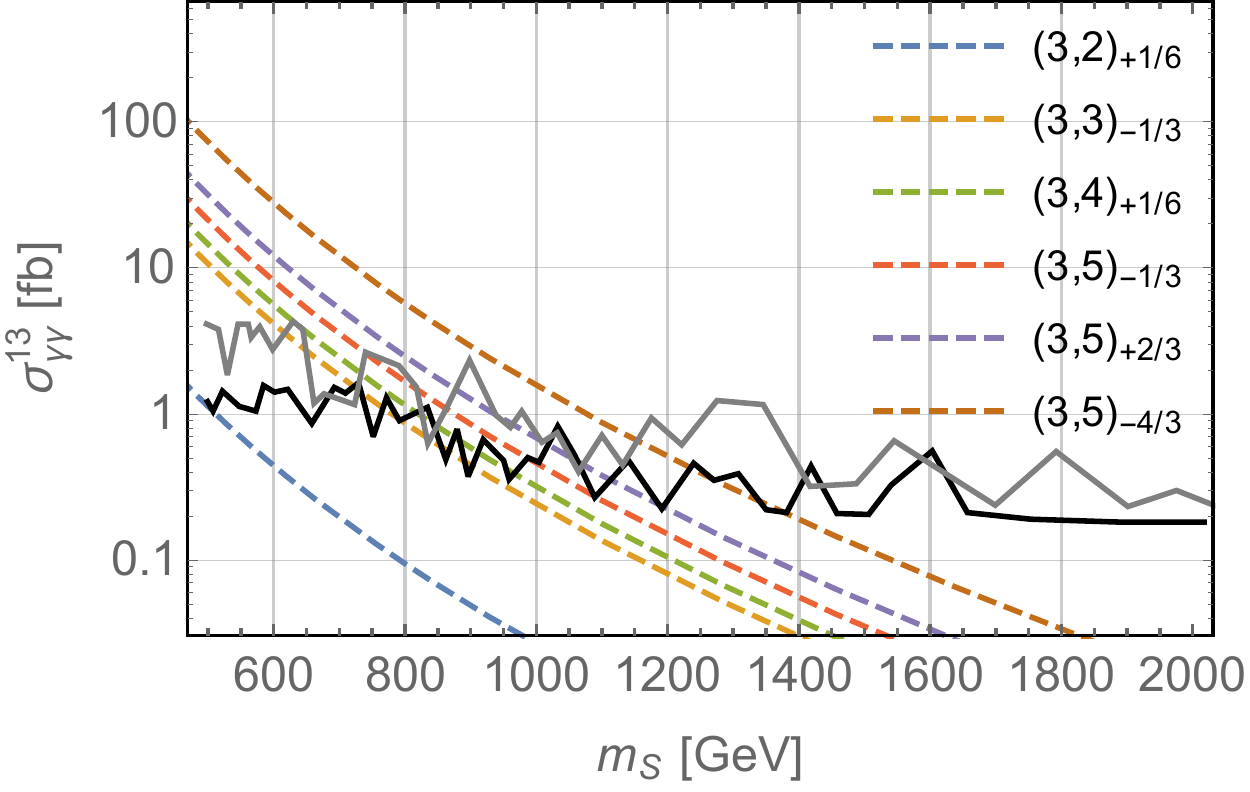}\label{fig:nonsing}
  }
  \caption{ \label{fig:ratios_diphoton} \it Diphoton signals for color-triplets with charges $\abs{Q}_{\rm high}=4/3,5/3,7/3,8/3,10/3$, at 13~TeV LHC. Line color represents a specific $\abs{Q}_{\rm high}$. The experimental upper bounds from ATLAS~\cite{ATLAS:2016eeo} (CMS~\cite{Khachatryan:2016yec}) are presented in black (gray). (a) $SU(2)_W$ singlets.  (b) High $SU(2)_W$ representations.}
\end{figure}

Interestingly, if $X$ transforms in a large $SU(2)_W$ representation, its total width can be much larger than its partial width into $gg$. This can deplete the various $S$ signals, in particular the $S\to\gamma\gamma$ one. We demonstrate this effect in Fig.~\ref{fig:ratios_diphoton}, where we show, for a given charge, the differences between the diphoton signal of an $SU(2)_W$ singlet to the one obtained from the highest $SU(2)_W$ representation listed in Table~\ref{tab:reps}. For the same charge $Q$ we  notice a dependence on the $SU(2)_W$ representation.
\begin{table}[h!]
\caption{ \it The ratio between the diphoton signals in the singlet and high-representation cases. The depletion of the signal for high representations is clearly seen for the various charges. The numbers are given for $m_S=800$~GeV and vary only little due to RGE effects.}
\label{tab:AAratios}
\begin{center}
\begin{tabular}{cc} \hline\hline
$Q_{\rm high/low}$ & $\sigma_{\gamma\gamma}$ ratio \\[2pt] \hline\hline
$+2/3$ & 1.0 \\
$-4/3$ & 1.6 \\
$+5/3$ & 2.4 \\
$-7/3$ & 3.4 \\
$+8/3$ & 2.7 \\
$-10/3$ & 1.4 \\
\hline\hline
\end{tabular}
\end{center}
\end{table}

The experimental upper bounds on $\sigma_{\gamma\gamma}$ at 13 TeV translate into a lower bound on $m_S$ and, consequently, on $m_X$. These bounds are effective: in fact, for $SU(2)_W$ singlets the bound is stronger than the bound from LHC direct continuum pair production searches in a large region of the parameter space. For instance, as discussed in the previous section, there are only very week bounds for an $X_{+5/3}$ state from direct continuum pair production searches, while the search for diphoton resonance gives $m_{X_{5/3}}\gtrsim600$~GeV.

For higher $SU(2)_W$ representations, the bound state limits can be weaker than the ones from direct continuum searches, but have the advantage of being less model dependent. Consider, for example, the quartet $X(3,4)_{+1/6}$. As discussed in the previous section, the lower bound on $m_X$ is very model dependent. It is around 800 GeV for decays into a leptoquark involving $e$ or $\mu$, but can be very weak for fully hadronic decays and reasonable mass splitting. Diphoton resonance searches set a solid bound of 450 GeV which is independent of these details of the model. Similar statements can be made for other high $SU(2)_W$ representations.

\subsection{Distinct features of a bound state composed of $SU(2)_W$-non-singlet constituents}
If an X-onium $S$ involves $X$ that is an $SU(2)_W$-non-singlet, then it might exhibit two features that would clearly distinguish it from the $SU(2)_W$-singlet case: a large branching ratio into $W^+W^-$ and an apparent large width. In this subsection we explain these two features.

{\bf Large BR($\mathbf{{S\to W^+W^-}}$)}: 
Observation of any diboson decay mode of $S$ -- $\gamma\gamma$, $W^+W^-$, $ZZ$, $Z\gamma$ -- will help to close in on the representation of $X$.
Our main focus is on cases where the $S\to W^+W^-$ decay rate is large. For the sake of concreteness, we examine
whether $R_{W W/\gamma\gamma}\geq10$ is possible. Tab.~\ref{tab:VVratios} shows five candidates. We list them by the order of the lower bound on their mass from diphoton searches:
\begin{itemize}
\item $(3,2)_{+1/6}$, with $m_S\gtrsim500$~GeV.
\item $(3,3)_{-1/3}$, with $m_S\gtrsim850$~GeV.
\item $(3,4)_{+1/6}$, with $m_S\gtrsim900$~GeV.
\item $(3,5)_{-1/3}$, with $m_S\gtrsim1.1$~TeV.
\item $(3,5)_{+2/3}$, with $m_S\gtrsim1.2$~TeV.
\end{itemize}
We assume that all members of the $X$-multiplet are close enough in mass that they are observed as a single $X$-onium resonance. Another option would be separated signatures, in which, for example, a diphoton signal would come mainly from the $\abs{Q}_{\rm high}$ state, while the $W^+W^-$ signature arises mainly from the $\abs{m}_{\rm low}$ state/s, possibly at different mass.

We note again that $X\sim(3, 4)_{+1/6}$ can be as light as 450 GeV only if $X_{+5/3} $ is the lightest state and the mass splitting is large enough to let all the other states decay to it via three body decay. We further discuss this possibility in the next section, in the context of the second scenario we study.

{\bf Large apparent $\mathbf{\Gamma_S}$:} The mass splitting between members of an $SU(2)_W$ multiplet may cause an apparent large width in the X-onium diphoton signal.
To this end, it is important that the contribution to the diphoton events is not completely dominated by a single member of the multiplet. However, since the contribution of a particle of charge $Q$ to the diphoton signal is proportional to $Q^4$, a single member dominance is the case more often than not. For example, for the $(3,2)_{-5/6}$ multiplet, the contribution of the $Q =-4/3$ particle is 256 times larger than that of the $Q =-1/3$ particle. From the representations in Tab.~\ref{tab:reps}, only two could result in an apparent large diphoton width:
\begin{itemize}
\item $(3, 4)_{+1/6}$, with $\sigma_{\gamma\gamma}^{-4/3}/\sigma_{\gamma\gamma}^{+5/3}\simeq0.40$.
\item $(3, 5)_{-1/3}$, with $\sigma_{\gamma\gamma}^{+5/3}/\sigma_{\gamma\gamma}^{-7/3}\simeq0.26$.
\end{itemize}

The mass splitting between two extreme bound states of an $SU(2)_W$ $n$-tuplet is $\Delta m_S\simeq-\lambda_{XH}^\prime (n-1)v^2/(2m_S)$. Therefore, a quartic coupling of size
\beqa
\abs{\lambda_{XH}^\prime}=\frac{1}{50(n-1)}\frac{m_S^2}{v^2}\,,
\eeqa
would saturate an estimated $1\%$ mass resolution of the diphoton signal (see, {\it e.g.}~\cite{ATLAS:2016eeo}). Such a small quartic coupling is allowed by EWPM and has no observed impact on Higgs couplings. Note that in order to understand whether the whole multiplet contributes to the resonance, or just the lightest member, one needs to make sure that the $W$-mediated decays within the multiplet, $X_m\to X_{m\pm1}W^{\mp(*)}$ (Eqs.~\eqref{eq:xwll} and~\eqref{eq:xwpi}), are not faster than the decay rate of $S$. This condition is generally satisfied below the $m_W$ threshold.

\subsection{Back to our model examples}
Let us now describe the phenomenology of the QCD bound state for our two benchmark scenarios of Sec.~\ref{sec:cases}.

\subsubsection{Degenerate $SU(2)_W$-quartet}
In this scenario with $m_X=520$~GeV, the bound state has a mass $m_S=1036$~GeV, with possible small splitting between the various $S_Q$ states. It exhibits the following features:
\begin{itemize}
\item {\bf $\gamma\gamma$}: Possible large apparent width in diphoton signals, with $\sigma_{\gamma\gamma}^{13}\simeq0.25$~fb.
\item {\bf $gg$}: Possible large apparent width in dijet signals, with $R_{gg/\gamma\gamma}\simeq11$.
\item {\bf $W^+W^-$}: Large $W^+W^-$ signal, with $R_{WW/\gamma\gamma}\simeq25$.
\item {\bf $ZZ$}: Enhanced $ZZ$ signal, with $R_{ZZ/\gamma\gamma}\simeq11$.
\end{itemize}
In particular, a discovery of $S$ with $m_S$ slightly above TeV is, in this case, within the reach of upcoming diphoton searches.

\subsubsection{Non-degenerate $SU(2)_W$-quartet}

This is an example in which the bound state search is more powerful than the direct searches of $X_Q$ due to the lack of sensitivity for the three body final state $Wjj$ which would allow quartet as light as 250 GeV. Diphoton searches for $S_Q$ exclude $m_S\leq900$~GeV, which corresponds to $m_X\lesssim450$~GeV. At the 13 TeV with increased luminosity we expect a resonance which exhibits the following features:
\begin{itemize}
\item {\bf $\gamma\gamma$}: Possibly two resolved diphoton resonances, with a total diphoton signal $\sigma_{\gamma\gamma}^{13}\simeq 0.58$~fb.
\item {\bf $gg$}: Wide dijet signal, with $R_{gg/\gamma\gamma}\simeq11$.
\item {\bf $W^+W^-$}: Large $W^+W^-$ signal, with $R_{WW/\gamma\gamma}\simeq25$.
\item {\bf $ZZ$}: Enhanced $ZZ$ signal, with $R_{ZZ/\gamma\gamma}\simeq11$.
\end{itemize}

\section{Summary and conclusions}\label{sec:concs}
The LHC search for new physics at or below the TeV scale is far from complete, even for strongly interacting particles. New particles might have surprising features, different from those predicted by the commonly studied extensions of the standard model. We studied the phenomenology of color-triplet scalar particles transforming in non-trivial representation of $SU(2)_W$ and potentially carrying exotic EM charges. Our main results are as follows.
\begin{itemize}
\item Color-triplet scalars ($X$), transforming in exotic representations of $SU(2)_W$ with masses at a few hundred GeV, are far from being experimentally excluded.
\item Depending on the electromagnetic charges of such colored scalars, their dominant decay modes could be into three or four body final states. Some of these decay topologies, in particular the $W^\pm jj$ one, are essentially unexplored by current analyses.
\item In large parts of the parameter space, $XX^\dagger$ for exotic $X$ would form a QCD-bound state ($S$). It is easy to find examples where the observation of di-electroweak boson ({\it e.g.} diphoton) resonance at $m_S$ will precede the direct discovery of $X$.
\item If $X$ is an $SU(2)_W$-non-singlet, the phenomenology of $S$ might involve intriguing features, such as $WW$ resonance at the same invariant mass as the diphoton resonance or somewhat removed from it, and a large apparent width for $S$.
\end{itemize}

\acknowledgments
We thank Daniel Aloni, Liron Barak, Jared Evans, Zhen Liu,  Adam Martin, Gilad Perez and Tomer Volansky for useful discussions.  YN is the Amos de-Shalit chair of theoretical physics.
This research is supported by the I-CORE program of the Planning and Budgeting Committee and the Israel Science Foundation (grant number 1937/12). YN is supported by grants from the Israel Science Foundation (grant number 394/16) and from the United States-Israel Binational Science Foundation (BSF), Jerusalem, Israel (grant number 2014230). The work of CF was carried out in part at the Aspen Center for Physics, which is supported by National Science Foundation grant PHY-1066293.

\appendix

\section{ $\left(W^+jj\right)\left(W^-jj\right)$ final state}\label{app:3body}

A dedicated search for the three body decay topology $Wjj$ has not been performed by the experimental collaborations.
There are, however, a few analyses which are potentially sensitive to this final state. As detailed in section~\ref{sec:direct}, we simulated our signal in MC simulation and compared between the efficiencies of these analyses for our $\left(W^+jj\right)\left(W^-jj\right)$ signal and for the topologies that originally served as benchmark models. We stress that the limits obtained in this way should be taken as indicative of the sensitivity of certain searches to our final state, rather than as a complete recast of the analyses.

{\bf The search for exotic vector-like fermions decaying into $ Wj $~\cite{Aad:2015tba}:} The bounds from this search are presented in Fig.~\ref{fig:direct} as they are found to be the most sensitive ones.
We find that the selection efficiencies of our signal and the targeted topology $\left(W^+j\right)\left(W^-j\right)$ are comparable. Yet, the binned analysis performed by the collaborations relies on the mass reconstruction of the parent fermion. Therefore our signal, originating from a three-body decay, suffers from a broadening of the $m_{Wj} $ distribution. We conservatively estimate this reduction to be between $30\%$ and $50\%$. Under this assumption, this search does not rule out the existence of a light quartet with all components cascading down to $ Wjj$, except at a small mass window between $375-440$~GeV.
At the Tevatron both CDF and D0~\cite{Aaltonen:2011tq,Abazov:2011vy} performed a similar search for fourth generation quarks in the mass range 200-500 GeV. Assuming similar reduction in the efficiency, these searches give the best sensitivity at the low mass ranges. They exclude, for example, an $SU(2)_W$ quartet below 250~GeV.

{\bf The search for supersymmetric multi-jet with $0/1/2$ leptons:}
\begin{itemize}
\item Fully hadronic:
Almost half of the events of our signal are purely hadronic. However, multi-jet searches suffer from the large QCD background and do not exclude a scalar color-triplet, even when taking into account the high multiplicity of a quartet.
(These searches are more effective for gluinos, which have a significantly higher cross section~\cite{Borschensky:2014cia}.)
\item One lepton: The final state contains a single lepton, jets and missing energy, which is common to many SUSY scenarios.
In particular, the ATLAS search of Ref.~\cite{Aad:2015mia} targets, among others, the double production of first and second generation squarks $ \tilde q$  decaying into $ W^\pm j \chi_0$ via on-shell chargino. An important parameter for this signal is $x=\Delta m(\chi^+,\chi^0)/\Delta m(\tilde q,\chi^0)$. For $x=1$ the squarks and chargino are degenerate, while for $x=0$ the chargino and neutralino are degenerate. Originating from a three body decay, the kinematics of the $W$'s in our signal resemble more the low $x$ case.
Taking $x=0.2$ as representative of the low $x$ region, we find that the efficiency of our signal is lower by $50\%$ than the one of the targeted signal. This reduction originates from the lower missing energy which, however, is partially compensated by the enlarged jet activity. Taking $x=0.8$ as representative of the high $x$ region, we find that the efficiency reduction becomes less than $10\%$. The similar analysis at 13~TeV has lower sensitivity as it typically targets higher masses~\cite{ATLAS:2016lsr,ATLAS:2016kts}.

\item Two leptons: Searches for a final state containing two leptons, missing energy and jets have a potentially similar reach, but pay a higher price in the leptonic branching ratio of the $W$ bosons. Therefore, they do not provide the best limits on our signal.
\end{itemize}

{\bf The search for first or second generation leptoquarks:} The LQ searches typically suffers from a $25\%$ reduction in the efficiency for our signal. This, together with the small leptonic $W$ branching ratios, yield bounds that are insignificant. We note that a mixed $(e^\pm j)(\mu^\mp j)$ search, which is currently not done by the collaborations, may have better sensitivity due to lower expected background.

{\bf Searches for various states containing $b$ jets:}
\begin{itemize}
\item The CMS 7  and 8~TeV analyses~\cite{Chatrchyan:2012vu,Khachatryan:2015oba} search for heavy top-like quark $(t^\prime)$ decaying to $Wb$ final state. These searches might be sensitive to a $Wbj$ topology. Yet, as previously discussed, the $t'$ mass reconstruction weakens the reach of this search to the $Wbj$ topology. We again estimate this reduction to be between $30\%$ and $50\%$ and show the resulting bounds in Fig.~\ref{fig:direct}. The same is done for the heavy bottom-like quark searches~\cite{Khachatryan:2016iqn,CMS:2016gvu,ATLAS:2016kjm}.
\item The CMS RPV-SUSY search~\cite{Khachatryan:2016iqn} for $ \tilde b \rightarrow t j$, where $\tilde b$ is the bottom squark, could have some sensitivity to $Wbj$ topology. However, it requires the reconstruction of $t$ quarks which reduces significantly the sensitivity to our signal.
\item SUSY stop searches, {\it e.g.}~\cite{Aad:2014kra}, look for a single lepton, missing energy and $b$-jets final state. We find these searches to be less sensitive than the heavy quark searches, as in the SUSY multi-jet searches with $1$ lepton.
\end{itemize}
We conclude that the $Wjj$ decay mode is presently poorly constrained, irrespective of the flavor of the jets in the final state.

{\bf Precision cross-section measurements:}
Precision measurement of the $t\bar t$ and $W^+W^-$ cross sections might probe best the low mass region of a $\left(W^+jj\right)\left(W^-jj\right)$ signal. However, for $m_X\geq250$~GeV we find that these are not sensitive even at multiplicity as high as $n=5$; the argument goes as follows.
We consider the NNLO-NNLL $t \bar t$ production cross section (see~\cite{Czakon:2011xx} and references therein), with $m_t = 172.5$ GeV, and combine scale uncertainty and the uncertainty associated with variations of the PDF and $\alpha_s$ (see~\cite{Botje:2011sn,Martin:2009bu,Gao:2013xoa,Ball:2012cx}). At $m_X=250$ GeV, the production cross section for a quintuplet is below the theoretical uncertainty, assuming the efficiency of the $\bar t t$ search to be $50 \% $ smaller than the efficiency for the  $\bar t t$ sample itself. This is a plausible estimate in the case of the $Wbj$ topology, and a conservative one for the $Wjj$ topology, even if we allow a large mistagging rate. Therefore, a quintuplet at~250 GeV is not constrained by the $ t \bar t $ measurements. As for the $W^+W^-$ cross section measurements, the relevant analyses veto on $N_j\geq1$. Since our signal contains many jets in the final state, it would not contribute significantly to these measurements.

\section{ $\left(W^+W^+jj\right)\left(W^-W^-jj\right)$ final state}\label{app:4body}

There  are no dedicated searches for the four body $WWjj$ decay mode, but other searches are potentially sensitive to it.
For the fully hadronic final states and for the ones containing only one or two leptons, conclusions similar to those made for the $Wjj$ decay mode hold. However, for this topology, the most promising search strategy is to look for multilepton final states. The low SM background compensates for the branching ratio suppression of four $W$'s decaying leptonically.

We analyze the RPV multilepton CMS search~\cite{Khachatryan:2016iqn,CMS:2016gvu} which does not rely on any missing energy cut. This analysis contains many exclusive signal regions, depending on the number of leptons, the presence of hadronically decaying $\tau$, the presence of $b$ jets, and the number of opposite-sign-same-flavor (OSSF) lepton pairs. We consider the low background regions, with four leptons, zero hadronic $\tau$'s and $1$ pair of OSSF leptons, summing over all $S_T$ bins. To be conservative, we allow the number of background events to fluctuate up by $95\%$ C.L. and the number of signal events to fluctuate down by $95\%$ C.L., assuming Poisson statistics. We take $N_{\rm sig}=\mathcal{L}\sigma\epsilon{\rm BR}_{4W\to4\ell,1OSSF}$ with very high efficiency $\epsilon=80\%-90\%$.

A somewhat stronger bound comes from the ATLAS analyses of Ref.~\cite{ATLAS:2016kjm}. For this, we consider the two overlapping signal regions, SR3L1 and SR0b1, with the corresponding bounds of $\sigma_{\rm SR3L1}\leq0.59$~fb and $\sigma_{\rm SR0b1}\leq0.37$~fb, set at $95\%$ C.L..  (For the exact description of these signal regions we refer the reader to Ref.~\cite{ATLAS:2016kjm}.) Since this search was specifically designed to be applicable to any SUSY RPV scenario, we assume the efficiency for our signal to be similar to the one quoted. We therefore use $\epsilon=2\%-5\%$.  The resulting limits are presented in Fig.~\ref{fig:direct}.

\section{ $\left(\ell^\pm t\right)\left(\ell^\mp\bar t\right)$ final state}\label{app:ellt}

Similar to the four-body decays, $X\to t\ell^\pm$ decay would be captured by the multi-lepton searches aiming at RPV SUSY signals. For this signature we estimate the reach of the CMS 8~TeV search~\cite{CMS:2016gvu} in the signal region with four leptons, zero hadronic taus, one pair of OSSF leptons and one tagged $b$-jet. As before, we allow upward fluctuation of the background and downward fluctuation of the signal, both within $95\%$ C.L.. Assuming efficiency of $60\%-80\%$, we find an excluded cross section of $\sigma^8_{XX^\dagger}\leq5-6.6$~fb. The resulting bounds as a function of $m_X$ are presented in Fig.~\ref{fig:direct}.

We note that the 13~TeV analysis of CMS~\cite{Khachatryan:2016iqn} veto $b$-jets, while the ATLAS 13~TeV analysis~\cite{ATLAS:2016kjm} uses large jet multiplicity $\left(N_j\geq6\right)$ and relatively large missing energy $\left(E_{\rm T}^{\rm miss}\geq200\right)$~GeV. Both of these searches are therefore less sensitive to our signal in this case.

\section{Running of gauge coupling constants}\label{app:RGE}

At one loop,
\beqa
\alpha(\mu)^{-1}=\alpha(\mu_0)^{-1}+\frac{\left(b^{\rm SM}+b^X\right)}{2\pi}\log\left(\frac{\mu}{\mu_0}\right)
\eeqa
with (we use the common GUT inspired definition $g_1=\sqrt{3/5}g^\prime$)
\beqa
b_1^{\rm SM}&&=-\frac{41}{10}\,,\;\;\;
b_2^{\rm SM}=\frac{19}{6}\,,\;\;\;
b_3^{\rm SM}=7\,,\no\\
b_1^X&&=-\frac{1}{5}RnY^2\,,\no\\
b_{2}^X&&=-\frac{1}{3}RC(n)=-\frac{1}{36}Rn(n^2-1)\,,\no\\
b_{3}^X&&=-\frac{1}{3}nC(R)=
  \begin{cases}
 -\frac{1}{6}n & {\rm for~}R=3 \\
 -\frac{5}{6}n & {\rm for~}R=6
  \end{cases}\,,
\eeqa
where $C(n)$ [$C(R)$] is the Casimir of the $n$ [$R$] representation of $SU(2)$ [$SU(3)$]. For numerical evaluation, we use, at $m_Z=91.1876$ GeV~\cite{Agashe:2014kda}
\beqa
\alpha_{\rm EM}(m_Z)&&=127.94^{-1}\,,\;\;\;
s_W^2(m_Z)=0.22333\,,\;\;\;\alpha_3(m_Z)=0.1185\no\\
\alpha_1&&=\frac{5}{3c_W^2}\alpha_{\rm EM}\,,\;\;\;
\alpha_2=\frac{1}{s_W^2}\alpha_{\rm EM}\,,
\eeqa
and $m_W=80.385$\,GeV.

\section{Effective operators}\label{app:EFTX}

Consider a scalar $X(R,n)_Y$ with the Lagrangian given in Eqs. (\ref{eq:lagX}) and (\ref{eq:Spotential}).
The impact of $X$ on the SM fields is mainly captured by the dimension six operators, generated at one-loop order upon integration out of $X$.
Refs.~ \cite{Henning:2014wua,Chiang:2015ura} compute the Wilson coefficient of these effective interactions for a general scalar. We present their results in Tab.~\ref{tab:dimsix}. Note that even though this list is not completely independent when the Higgs and gauge bosons equations of motion are considered, we find it convenient for our purposes to determine the oblique parameters and the Higgs couplings, as long as no redundancy is used when considering physical parameters.
\begin{table}[h!]
\caption{\it Dimension six operators generated by integrating out a scalar $X(R,n)_Y$.}
\label{tab:dimsix}
\begin{center}
\begin{tabular}{lllc} \hline\hline
\rule{0pt}{1.2em}%
operator & & 1-loop Wilson coefficient & physical importance  \\[2pt] \hline\hline
\rule{0pt}{1.2em}%
$\mathcal{O}_H$ & $\frac{1}{2}\left(\partial_\mu\abs{H}^2\right)^2$ & $\frac{R}{16\pi^2m_X^2}\frac{n\lambda_{XH}^2}{6}$ & $\delta c_W$, $\delta c_Z$ and $\delta c_f$ \\
$\mathcal{O}_T$ & $\frac{1}{2}\left(H^\dagger\overleftrightarrow D_\mu H\right)^2$ & $\frac{R}{16\pi^2m_X^2}\frac{n(n^2-1)\lambda_{XH}^{\prime 2}}{288}$ & $T$ parameter and $\delta c_W$ \\
$\mathcal{O}_{WW}$ & $g^2H^\dagger HW_{\mu\nu}^aW^{a\mu\nu}$ & $\frac{R}{16\pi^2m_X^2}\frac{n\left(n^2-1\right)\lambda_{XH}}{144}$ & $\delta c_\gamma$ \\
$\mathcal{O}_{BB}$ & $g^{\prime 2}H^\dagger HB_{\mu\nu}B^{\mu\nu}$ & $\frac{R}{16\pi^2m_X^2}\frac{nY^2\lambda_{XH}}{12}$ & $\delta c_\gamma$ \\
$\mathcal{O}_{WB}$ & $2gg^\prime \left(H^\dagger\tau^aH\right)\left(W_{\mu\nu}^aB^{\mu\nu}\right)$ & $\frac{R}{16\pi^2m_X^2}\frac{n\left(n^2-1\right)Y\lambda_{XH}^\prime}{144}$ & $S$ parameter, $\delta c_\gamma$ and $\delta c_W$ \\
$\mathcal{O}_{GG}$ & $g_s^2H^\dagger HG_{\mu\nu}^aG^{a\mu\nu}$ & $\frac{1}{16\pi^2m_X^2}\frac{n\lambda_{XH}C(R)}{12}$ & $\delta c_g$ \\
\hline\rule{0pt}{1.0em}%
$\mathcal{O}_{2W}$ & $-\frac{1}{2}\left(D_\mu W^a_{\mu\nu}\right)^2$ & $\frac{R}{16\pi^2m_X^2}\frac{n(n^2-1)g^2}{360}$ & $b_2$ and $W$ parameter \\
$\mathcal{O}_{2B}$ & $-\frac{1}{2}\left(\partial_\mu B_{\mu\nu}\right)^2$ & $\frac{R}{16\pi^2m_X^2}\frac{nY^2g^{\prime2}}{30}$ & $b_1$ and $Y$ parameter \\
$\mathcal{O}_{2G}$ & $-\frac{1}{2}\left(D_\mu G^\lambda_{\mu\nu}\right)^2$ & $\frac{C(R)}{16\pi^2m_X^2}\frac{g_3^2}{30}$ & $b_3$ \\
\hline\hline
\end{tabular}
\end{center}
\end{table}

\section{Oblique parameters}
A scalar $X(R,n)_Y$ alters the vacuum polarization amplitudes of the EW gauge fields. These effects are conveniently parameterized by the oblique parameters $S,T$ and $U$~\cite{Peskin:1991sw} and $V,X,Y$ and $W$ (for a review see~\cite{Barbieri:2004qk}). The leading contributions to the oblique parameters read
\beqa
T=&&\frac{v^2}{\alpha}c_T=\left(\frac{v^2}{4608\pi^2\alpha}\right)\left(\frac{\lambda_{XH}^{\prime 2}}{m_X^2}\right)Rn(n^2-1)\,,\no\\
S=&&16\pi v^2c_{WB}=\left(\frac{v^2}{144\pi}\right)\left(\frac{\lambda_{XH}^\prime}{m_X^2}\right)Rn(n^2-1)Y\,,\no\\
Y=&&2m_W^2c_{2B}=\frac{g^{\prime2}}{240\pi^2}\frac{m_W^2}{m_X^2}RnY^2\,,\no\\
W=&&2m_W^2c_{2W}=\frac{g^2}{2880\pi^2}\frac{m_W^2}{m_X^2}Rn\left(n^2-1\right)\,,\no\\
\eeqa

\section{Higgs couplings}\label{app:Higgs}

The quartic scalar couplings $\lambda_{XH}$ and $\lambda^\prime_{XH}$ modify the light Higgs couplings from their SM values. For an $X(R,n)_Y$ representation, these modifications read
\beqa
\delta c_\gamma&&=4\pi^2v^2\left(c_{BB}+c_{WW}-c_{WB}\right)=\frac{v^2}{144 m_X^2}nR\left\{\left[\frac{n^2-1}{2}+3Y^2\right]\lambda-\frac{\left(n^2-1\right)Y\lambda^\prime}{2}\right\}\,,\no\\
\delta c_g&&=48\pi^2v^2c_{GG}=\frac{v^2\lambda_{XH}}{4m_X^2}nC(R)\,,\no\\
\delta c_W&&=-c_H\frac{v^2}{2}+\frac{2c_W^2v^2}{c_W^2-s_W^2}c_T-\frac{16\pi\alpha v^2}{c_W^2-s_W^2}c_{WB}\no\\
&&=\frac{v^2}{12m_X^2}nR\left[-\frac{\lambda_{XH}^2}{16\pi^2}+\frac{c_W^2\left(n^2-1\right)\lambda_{XH}^{\prime2}}{96(c_W^2-s_W^2)}
+\frac{\alpha Y\left(n^2-1\right)\lambda_{XH}^\prime}{6\pi(c_W^2-s_W^2)}\right]\,,\no\\
\delta c_Z&&=-c_H\frac{v^2}{2}=-\frac{v^2\lambda_{XH}^2}{192\pi^2m_X^2}nR\,,\no\\
\delta c_f&&=-c_H\frac{v^2}{2}=-\frac{v^2\lambda_{XH}^2}{192\pi^2m_X^2}nR\,.
\eeqa
The $hgg$ and $h\gamma\gamma$ couplings are computed using the Higgs effective low energy theory~\cite{Kniehl:1995tn}:
\beqa
\delta c_\gamma&&=\frac{R}{24}\sum_QQ^2\frac{v\partial\log M_Q}{\partial v}\,,\no\\
\delta c_g&&=\frac{C(R)}{2}\sum_Q\frac{v\partial\log M_Q}{\partial v}\,,
\eeqa
where
\beqa
M_Q^2=m_X^2+\left(\lambda_{XH}-\frac{\lambda_{XH}^\prime Q}{2}\right)\frac{v^2}{2}\,.
\eeqa
Other couplings are computed by their definition in terms of the Wilson coefficients, for which we use the results of Refs.~\cite{Agashe:2014kda,Falkowski:2015wza}.

\begin{figure}[h!]
	\begin{center}
	\includegraphics[height=2in]{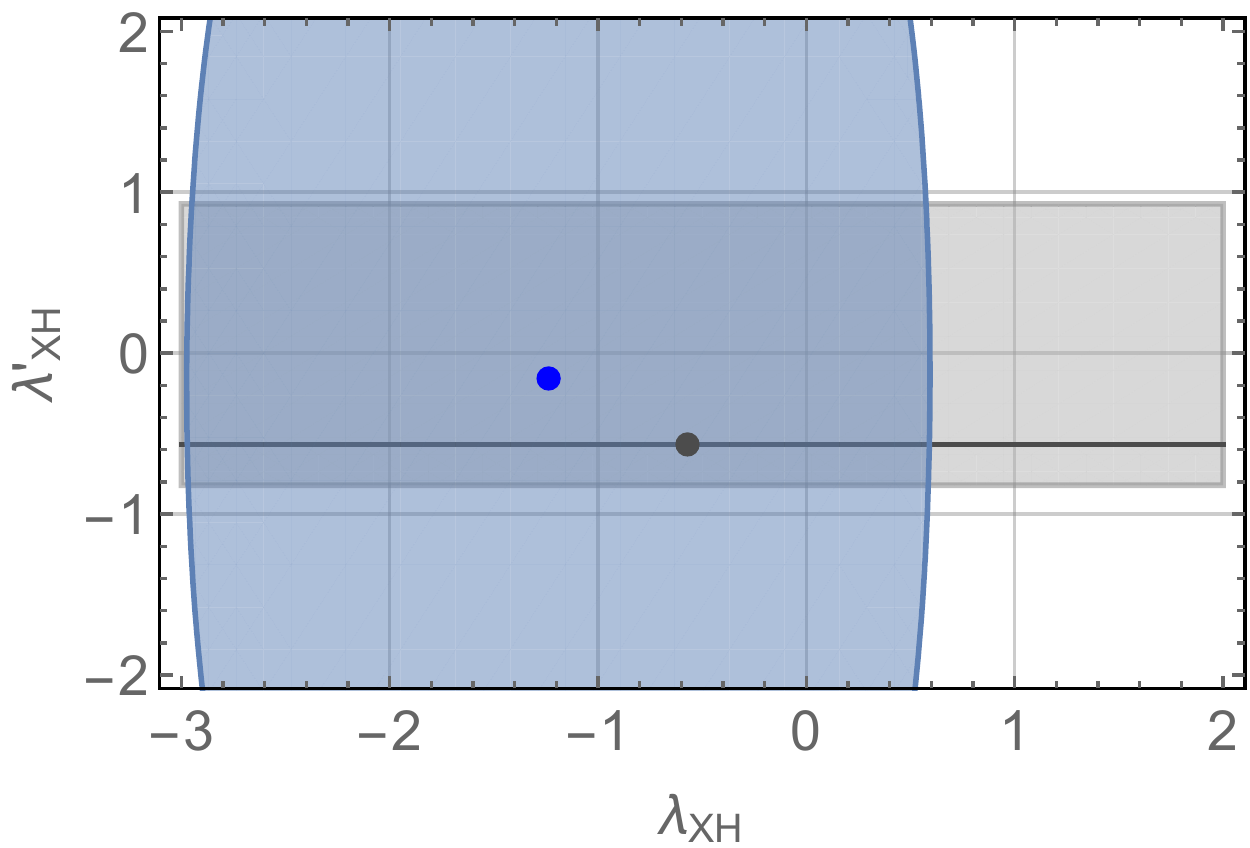}
\caption{\it Higgs decay (Blue) and EWPM (Gray) constraints on the quartic couplings $\lambda_{XH}$ and $\lambda^\prime_{XH}$ at $95\%$ C.L. for $X\sim(3,4)_{+1/6}$ at $m_X=450$~GeV. Blue point is the best fit value from the Higgs data. Black line is the best fit value from EWPM.}
\label{fig:expresults}
	\end{center}
\end{figure}

For our numerical results we use table~14 of~\cite{Higgs} with $B_{\rm BSM}=0$. We take as a concrete example the case of $X\sim(3,4)_{+1/6}$. The exact results, including EWPM constraints, are shown in Fig.~\ref{fig:expresults}. The constraints on $\lambda_{XH}$ and $\lambda^\prime_{XH}$ are rather mild and do not affect our conclusions.

\section{Quartic coupling running}\label{app:RGEquartic}
In this appendix we obtain the one loop $\beta$ function for the four quartic couplings of the scalar potential, where
\beqa
\frac{{\rm d}\lambda}{{\rm d}\log \mu}=\beta_\lambda\,.
\eeqa
We use the normalization $\lambda_H=m_h^2/v^2$ for the SM Higgs quartic coupling. Other quartic couplings are defined in Eq.~\eqref{eq:Spotential}. EW corrections of the order $g^2,g^{\prime 2}$ are neglected. The $\lambda_H$ RGE is given by
\beqa
\beta_{\lambda_H}
&&=\frac{1}{16\pi^2}\left[12\lambda_H^2+2Rn\lambda_{XH}^2+\frac{Rn(n^2-1)}{24}\lambda_{XH}^{\prime2}+\ldots\right]\,,
\eeqa
where $\ldots$ indicates other SM contributions (coming from, {\it e.g.}, the top-quark). The $\lambda_X$ RGE is given by
\beqa
\beta_{\lambda_X}&&=\frac{1}{16\pi^2}\left[(2Rn+8)\lambda_X^2+4\lambda_{XH}^2
+\frac{13}{3}g_3^4-16g_3^2\lambda_X\right]\,.
\eeqa
As for the mixed terms, we find:
\beqa
\beta_{\lambda_{XH}}&&=\frac{1}{16\pi^2}
\left[2(Rn+1)\lambda_X\lambda_{XH}+6\lambda_H\lambda_{XH}+4\lambda_{XH}^2
+\left(\frac{n^2-1}{4}\right)\lambda_{XH}^{\prime2}-8g_3^2\lambda_{XH}\right]\,,\no\\
\beta_{\lambda^\prime_{XH}}&&=\frac{1}{16\pi^2}\left[2\lambda_X\lambda^\prime_{XH}+2\lambda_H\lambda^\prime_{XH}+8\lambda_{XH}\lambda^\prime_{XH}-8g_3^2\lambda_{XH}^\prime\right]\,.
\eeqa

\section{$S\to VV$ decays}\label{app:rates}

For the $S\to VV$ decays, we use (see~\cite{Kats:2012ym} and references therein)
\beq
\Gamma(S_Q\to V_1V_2)=\frac{R}{8\pi(1+\delta_{V_1V_2})}\lambda^{1/2}(M_S^2,m_{V_1}^2,m_{V_2}^2)\frac{|\psi(0)|^2}{M_S^2}|{\overline{\cal M}}_{V_1V_2}|^2,
\eeq
where $\lambda[x,y,z]$ is defined below Eq.~\eqref{eq:jj}, and the squared amplitudes are given by~\cite{Drees:1993uw,Martin:2008sv}
\beqa
\abs{\mathcal{\overline M}_{gg}}^2=&&C_2(R)^2\left(16\pi^2\alpha_s^2\right)\,,\no\\
\abs{\mathcal{\overline M}_{\gamma\gamma}}^2=&&8e^4Q^4\,,\no\\
\abs{\mathcal{\overline M}_{ZZ}}^2=&&\frac{8e^4\left(m-Qs_W^2\right)^4}{s_W^4c_W^4}\left(1+\frac{0.1\tilde\lambda_{XH}^m}{\left(m-Qs_W^2\right)^2}+\frac{0.7(\tilde\lambda^m_{XH})^{2}}{\left(m-Qs_W^2\right)^4}\right)\,,\no\\
\abs{\mathcal{\overline M}_{Z\gamma}}^2=&&8\left(\frac{m-Qs_W^2}{s_Wc_W}\right)^2Q^2e^4\,,\no\\
\abs{\mathcal{\overline M}_{WW}}^2=&&\frac{e^4\left(n^2-1-4m^2\right)^2}{8s_W^4}\left(1+\frac{0.7\tilde\lambda_{XH}^m}{\left(n^2-1-4m^2\right)}+\frac{11(\tilde\lambda^m_{XH})^{2}}{\left(n^2-1-4m^2\right)^2}\right)\,,\no\\
\abs{\mathcal{\overline M}_{hh}}^2=&&\left[\tilde\lambda^m_{XH}-3\tilde\lambda^m_{XH}\left(\frac{m_h^2}{4m_X^2-m_h^2}\right)+(\tilde\lambda^m_{XH})^{2}\left(\frac{2v^2}{2m_X^2-m_h^2}\right)\right]^2\,.
\eeqa
Here $C_2(R)$ is the quadratic Casimir, with $C_2(3)=4/3$. The quartic coupling $\tilde\lambda_{XH}^m=\lambda_{XH}-(m/2)\lambda_{XH}^\prime$ (with $Q=m+Y$) changes the decay rates of $S_Q$ into $WW$ and $ZZ$ final states in a mild way, and generates $S_Q\to hh$ decays.



\begin{thebibliography}{99}

\bibitem{Dimopoulos:1981zb}
  S.~Dimopoulos and H.~Georgi,
  Nucl.\ Phys.\ B {\bf 193}, 150 (1981).

\bibitem{Kaplan:1983sm}
  D.~B.~Kaplan, H.~Georgi and S.~Dimopoulos,
  Phys.\ Lett.\ B {\bf 136}, 187 (1984).

\bibitem{Giudice:2013nak}
  G.~F.~Giudice,
  PoS EPS {\bf -HEP2013}, 163 (2013)
  [arXiv:1307.7879 [hep-ph]].

\bibitem{Kats:2012ym}
  Y.~Kats and M.~J.~Strassler,
  JHEP {\bf 1211}, 097 (2012)
  Erratum: [JHEP {\bf 1607}, 009 (2016)]
  [arXiv:1204.1119 [hep-ph]].

\bibitem{Kats:2016kuz}
  Y.~Kats and M.~J.~Strassler,
  JHEP {\bf 1605}, 092 (2016)
  Erratum: [JHEP {\bf 1607}, 044 (2016)]
  [arXiv:1602.08819 [hep-ph]].

\bibitem{Craig:2015yvw}
 N.~Craig,
  arXiv:1512.06819 [hep-ph].

\bibitem{Fox:2002bu}
  P.~J.~Fox, A.~E.~Nelson and N.~Weiner,
  JHEP {\bf 0208}, 035 (2002)
  [hep-ph/0206096].

\bibitem{Pati:1974yy}
  J.~C.~Pati and A.~Salam,
  Phys.\ Rev.\ D {\bf 10}, 275 (1974)
  Erratum: [Phys.\ Rev.\ D {\bf 11}, 703 (1975)].

\bibitem{Chacko:1998td}
  Z.~Chacko and R.~N.~Mohapatra,
  Phys.\ Rev.\ D {\bf 59}, 055004 (1999)
  [hep-ph/9802388].

\bibitem{Alloul:2013bka}
  A.~Alloul, N.~D.~Christensen, C.~Degrande, C.~Duhr and B.~Fuks,
  Comput.\ Phys.\ Commun.\  {\bf 185}, 2250 (2014)
  [arXiv:1310.1921 [hep-ph]].

\bibitem{Alwall:2011uj}
  J.~Alwall, M.~Herquet, F.~Maltoni, O.~Mattelaer and T.~Stelzer,
  JHEP {\bf 1106}, 128 (2011)
  [arXiv:1106.0522 [hep-ph]].

\bibitem{Sjostrand:2006za}
  T.~Sjostrand, S.~Mrenna and P.~Z.~Skands,
  JHEP {\bf 0605}, 026 (2006)
  [hep-ph/0603175].

\bibitem{Sjostrand:2007gs}
  T.~Sjostrand, S.~Mrenna and P.~Z.~Skands,
  Comput.\ Phys.\ Commun.\  {\bf 178}, 852 (2008)
  [arXiv:0710.3820 [hep-ph]].

\bibitem{deFavereau:2013fsa}
  J.~de Favereau {\it et al.} [DELPHES 3 Collaboration],
  JHEP {\bf 1402}, 057 (2014)
  [arXiv:1307.6346 [hep-ex]].

\bibitem{Beenakker:2010nq}
  W.~Beenakker, S.~Brensing, M.~Kramer, A.~Kulesza, E.~Laenen and I.~Niessen,
  JHEP {\bf 1008}, 098 (2010)
  [arXiv:1006.4771 [hep-ph]].

\bibitem{Kramer:2012bx}
  M.~Kramer, A.~Kulesza, R.~van der Leeuw, M.~Mangano, S.~Padhi, T.~Plehn and X.~Portell,
  arXiv:1206.2892 [hep-ph].

\bibitem{Borschensky:2014cia}
  C.~Borschensky, M.~Krämer, A.~Kulesza, M.~Mangano, S.~Padhi, T.~Plehn and X.~Portell,
  Eur.\ Phys.\ J.\ C {\bf 74}, no. 12, 3174 (2014)
  [arXiv:1407.5066 [hep-ph]].

\bibitem{Csaki:2015uza}
  C.~Csaki, E.~Kuflik, S.~Lombardo, O.~Slone and T.~Volansky,
  JHEP {\bf 1508}, 016 (2015)
  [arXiv:1505.00784 [hep-ph]].

\bibitem{Liu:2015bma}
  Z.~Liu and B.~Tweedie,
  JHEP {\bf 1506}, 042 (2015)
  [arXiv:1503.05923 [hep-ph]].

\bibitem{Khachatryan:2014lpa}
  V.~Khachatryan {\it et al.} [CMS Collaboration],
  Phys.\ Lett.\ B {\bf 747}, 98 (2015)
  [arXiv:1412.7706 [hep-ex]].

\bibitem{ATLAS:2012ds}
  G.~Aad {\it et al.} [ATLAS Collaboration],
  Eur.\ Phys.\ J.\ C {\bf 73}, no. 1, 2263 (2013)
  [arXiv:1210.4826 [hep-ex]].

\bibitem{Chatrchyan:2013izb}
  S.~Chatrchyan {\it et al.} [CMS Collaboration],
  Phys.\ Rev.\ Lett.\  {\bf 110}, no. 14, 141802 (2013)
  [arXiv:1302.0531 [hep-ex]].

\bibitem{ATLAS:084}
  The ATLAS collaboration [ATLAS Collaboration],
  ATLAS-CONF-2016-084.

\bibitem{Aad:2014aqa}
  G.~Aad {\it et al.} [ATLAS Collaboration],
  Phys.\ Rev.\ D {\bf 91}, no. 5, 052007 (2015)
  [arXiv:1407.1376 [hep-ex]].


\bibitem{ATLAS:2016yhq}
  The ATLAS collaboration [ATLAS Collaboration],
  ATLAS-CONF-2016-022.

\bibitem{Aad:2016kww}
  G.~Aad {\it et al.} [ATLAS Collaboration],
  JHEP {\bf 1606}, 067 (2016)
  [arXiv:1601.07453 [hep-ex]].

\bibitem{Khachatryan:2016iqn}
  V.~Khachatryan {\it et al.} [CMS Collaboration],
  arXiv:1606.08076 [hep-ex].

\bibitem{Chatrchyan:2013oba}
  S.~Chatrchyan {\it et al.} [CMS Collaboration],
  JHEP {\bf 1406}, 125 (2014)
  [arXiv:1311.5357 [hep-ex]].

\bibitem{Aaltonen:2011tq}
  T.~Aaltonen {\it et al.} [CDF Collaboration],
  Phys.\ Rev.\ Lett.\  {\bf 107}, 261801 (2011)
  [arXiv:1107.3875 [hep-ex]].

\bibitem{Abazov:2011vy}
  V.~M.~Abazov {\it et al.} [D0 Collaboration],
  Phys.\ Rev.\ Lett.\  {\bf 107}, 082001 (2011)
  [arXiv:1104.4522 [hep-ex]].

\bibitem{Aad:2015tba}
  G.~Aad {\it et al.} [ATLAS Collaboration],
  Phys.\ Rev.\ D {\bf 92}, no. 11, 112007 (2015)
  [arXiv:1509.04261 [hep-ex]].

\bibitem{Aaltonen:2011vr}
  T.~Aaltonen {\it et al.} [CDF Collaboration],
  Phys.\ Rev.\ Lett.\  {\bf 106}, 141803 (2011)
  [arXiv:1101.5728 [hep-ex]].

\bibitem{Aad:2015kqa}
  G.~Aad {\it et al.} [ATLAS Collaboration],
  JHEP {\bf 1508}, 105 (2015)
  [arXiv:1505.04306 [hep-ex]].

\bibitem{Chatrchyan:2012vu}
  S.~Chatrchyan {\it et al.} [CMS Collaboration],
  Phys.\ Lett.\ B {\bf 718}, 307 (2012)
  [arXiv:1209.0471 [hep-ex]].

\bibitem{CMS:2012ab}
  S.~Chatrchyan {\it et al.} [CMS Collaboration],
  Phys.\ Lett.\ B {\bf 716} (2012) 103
  [arXiv:1203.5410 [hep-ex]].

\bibitem{Aad:2015mba}
  G.~Aad {\it et al.} [ATLAS Collaboration],
  Phys.\ Rev.\ D {\bf 91} (2015) no.11,  112011
  [arXiv:1503.05425 [hep-ex]].

\bibitem{ATLAS:2016sno}
  The ATLAS collaboration [ATLAS Collaboration],
  ATLAS-CONF-2016-032.

\bibitem{Chatrchyan:2012yea}
  S.~Chatrchyan {\it et al.} [CMS Collaboration],
  JHEP {\bf 1205} (2012) 123
  [arXiv:1204.1088 [hep-ex]].

\bibitem{ATLAS:2016kjm}
  The ATLAS collaboration [ATLAS Collaboration],
  ATLAS-CONF-2016-037.

\bibitem{CMS:2016gvu}
  CMS Collaboration [CMS Collaboration],
  CMS-PAS-SUS-16-024.

\bibitem{Chatrchyan:2012vza}
  S.~Chatrchyan {\it et al.} [CMS Collaboration],
  Phys.\ Rev.\ D {\bf 86}, 052013 (2012)
  [arXiv:1207.5406 [hep-ex]].

\bibitem{Aad:2015caa}
  G.~Aad {\it et al.} [ATLAS Collaboration],
  Eur.\ Phys.\ J.\ C {\bf 76}, no. 1, 5 (2016)
  [arXiv:1508.04735 [hep-ex]].

\bibitem{Aad:2011ch}
  G.~Aad {\it et al.} [ATLAS Collaboration],
  Phys.\ Lett.\ B {\bf 709}, 158 (2012)
  Erratum: [Phys.\ Lett.\ B {\bf 711}, 442 (2012)]
  [arXiv:1112.4828 [hep-ex]].

\bibitem{Khachatryan:2015vaa}
  V.~Khachatryan {\it et al.} [CMS Collaboration],
  Phys.\ Rev.\ D {\bf 93}, no. 3, 032004 (2016)
  [arXiv:1509.03744 [hep-ex]].

\bibitem{Aaboud:2016qeg}
  M.~Aaboud {\it et al.} [ATLAS Collaboration],
  arXiv:1605.06035 [hep-ex].

\bibitem{CMS:2016imw}
  CMS Collaboration [CMS Collaboration],
  CMS-PAS-EXO-16-043.

\bibitem{ATLAS:2012aq}
  G.~Aad {\it et al.} [ATLAS Collaboration],
  Eur.\ Phys.\ J.\ C {\bf 72}, 2151 (2012)
  [arXiv:1203.3172 [hep-ex]].

\bibitem{Aaboud:2016zdn}
  M.~Aaboud {\it et al.} [ATLAS Collaboration],
  Eur.\ Phys.\ J.\ C {\bf 76}, no. 7, 392 (2016)
  [arXiv:1605.03814 [hep-ex]].

\bibitem{Aad:2014wea}
  G.~Aad {\it et al.} [ATLAS Collaboration],
  JHEP {\bf 1409}, 176 (2014)
  [arXiv:1405.7875 [hep-ex]].

\bibitem{Khachatryan:2015bsa}
  V.~Khachatryan {\it et al.} [CMS Collaboration],
  JHEP {\bf 1507}, 042 (2015)
  [arXiv:1503.09049 [hep-ex]].

\bibitem{Khachatryan:2014ura}
  V.~Khachatryan {\it et al.} [CMS Collaboration],
  Phys.\ Lett.\ B {\bf 739}, 229 (2014)
  [arXiv:1408.0806 [hep-ex]].

\bibitem{CMS:2016hsa}
  CMS Collaboration [CMS Collaboration],
  CMS-PAS-EXO-16-023.

\bibitem{Aaboud:2016lwz}
  M.~Aaboud {\it et al.} [ATLAS Collaboration],
  arXiv:1606.03903 [hep-ex].

\bibitem{Cirelli:2005uq}
  M.~Cirelli, N.~Fornengo and A.~Strumia,
  Nucl.\ Phys.\ B {\bf 753}, 178 (2006)
  [hep-ph/0512090].

\bibitem{Peskin:1991sw}
  M.~E.~Peskin and T.~Takeuchi,
  Phys.\ Rev.\ D {\bf 46}, 381 (1992).

\bibitem{Baak:2014ora}
  M.~Baak {\it et al.} [Gfitter Group Collaboration],
  Eur.\ Phys.\ J.\ C {\bf 74}, 3046 (2014)
  [arXiv:1407.3792 [hep-ph]].

\bibitem{Barbieri:2004qk}
  R.~Barbieri, A.~Pomarol, R.~Rattazzi and A.~Strumia,
  Nucl.\ Phys.\ B {\bf 703}, 127 (2004)
  [hep-ph/0405040].

\bibitem{Alves:2014cda}
  D.~S.~M.~Alves, J.~Galloway, J.~T.~Ruderman and J.~R.~Walsh,
  JHEP {\bf 1502}, 007 (2015)
  [arXiv:1410.6810 [hep-ph]].

\bibitem{Farina:2016rws}
  M.~Farina, G.~Panico, D.~Pappadopulo, J.~T.~Ruderman, R.~Torre and A.~Wulzer,
  arXiv:1609.08157 [hep-ph].

\bibitem{Goertz:2016iwa}
  F.~Goertz, A.~Katz, M.~Son and A.~Urbano,
  arXiv:1602.04801 [hep-ph].

\bibitem{Kahawala:2011pc}
  D.~Kahawala and Y.~Kats,
  JHEP {\bf 1109}, 099 (2011)
  [arXiv:1103.3503 [hep-ph]].

\bibitem{ATLAS:2016eeo}
  The ATLAS collaboration [ATLAS Collaboration],
  ATLAS-CONF-2016-059.

\bibitem{Khachatryan:2016yec}
  V.~Khachatryan {\it et al.} [CMS Collaboration],
  arXiv:1609.02507 [hep-ex].

\bibitem{Aad:2015mia}
  G.~Aad {\it et al.} [ATLAS Collaboration],
  JHEP {\bf 1504}, 116 (2015)
  [arXiv:1501.03555 [hep-ex]].

\bibitem{ATLAS:2016lsr}
  The ATLAS collaboration [ATLAS Collaboration],
  ATLAS-CONF-2016-054.

\bibitem{ATLAS:2016kts}
  The ATLAS collaboration [ATLAS Collaboration],
  ATLAS-CONF-2016-078.

\bibitem{Khachatryan:2015oba}
  V.~Khachatryan {\it et al.} [CMS Collaboration],
  Phys.\ Rev.\ D {\bf 93}, no. 1, 012003 (2016)
  [arXiv:1509.04177 [hep-ex]].

\bibitem{Aad:2014kra}
  G.~Aad {\it et al.} [ATLAS Collaboration],
  JHEP {\bf 1411}, 118 (2014)
  [arXiv:1407.0583 [hep-ex]].

\bibitem{Czakon:2011xx}
  M.~Czakon and A.~Mitov,
  Comput.\ Phys.\ Commun.\  {\bf 185}, 2930 (2014)
  [arXiv:1112.5675 [hep-ph]].

\bibitem{Botje:2011sn}
  M.~Botje {\it et al.},
  arXiv:1101.0538 [hep-ph].

\bibitem{Martin:2009bu}
  A.~D.~Martin, W.~J.~Stirling, R.~S.~Thorne and G.~Watt,
  Eur.\ Phys.\ J.\ C {\bf 64}, 653 (2009)
  [arXiv:0905.3531 [hep-ph]].

\bibitem{Gao:2013xoa}
  J.~Gao {\it et al.},
  Phys.\ Rev.\ D {\bf 89}, no. 3, 033009 (2014)
  [arXiv:1302.6246 [hep-ph]].

\bibitem{Ball:2012cx}
  R.~D.~Ball {\it et al.},
  Nucl.\ Phys.\ B {\bf 867}, 244 (2013)
  [arXiv:1207.1303 [hep-ph]].

\bibitem{Agashe:2014kda}
  K.~A.~Olive {\it et al.} [Particle Data Group Collaboration],
  Chin.\ Phys.\ C {\bf 38}, 090001 (2014).

\bibitem{Chiang:2015ura}
  C.~W.~Chiang and R.~Huo,
  JHEP {\bf 1509}, 152 (2015)
  [arXiv:1505.06334 [hep-ph]].

\bibitem{Henning:2014wua}
  B.~Henning, X.~Lu and H.~Murayama,
  JHEP {\bf 1601}, 023 (2016)
  [arXiv:1412.1837 [hep-ph]].

\bibitem{Kniehl:1995tn}
  B.~A.~Kniehl and M.~Spira,
  Z.\ Phys.\ C {\bf 69}, 77 (1995)
  [hep-ph/9505225].

\bibitem{Falkowski:2015wza}
  A.~Falkowski, B.~Fuks, K.~Mawatari, K.~Mimasu, F.~Riva and V.~sanz,
  Eur.\ Phys.\ J.\ C {\bf 75}, no. 12, 583 (2015)
  [arXiv:1508.05895 [hep-ph]].

\bibitem{Higgs}
  The ATLAS and CMS Collaborations,
  ATLAS-CONF-2015-044.

\bibitem{Drees:1993uw}
  M.~Drees and M.~M.~Nojiri,
  Phys.\ Rev.\ D {\bf 49}, 4595 (1994)
  [hep-ph/9312213].

\bibitem{Martin:2008sv}
  S.~P.~Martin,
  Phys.\ Rev.\ D {\bf 77}, 075002 (2008)
  [arXiv:0801.0237 [hep-ph]].


\end{thebibliography}
\end{document}